\documentclass[12pt]{article}
\usepackage{mymacros}
\begin{document}

\subheader{\begin{flushright}
\texttt{IFT-UAM/CSIC-23-95}
\end{flushright}}

\title{Entangled universes in dS wedge holography}

\author[a]{Sergio E. Aguilar-Gutierrez,}
\author[b]{Ayan K. Patra}
\author[b]{and Juan F. Pedraza}
\affiliation[a]{Institute for Theoretical Physics, KU Leuven, 3001 Leuven, Belgium}
\affiliation[b]{Instituto de Fisica Teorica UAM/CSIC, Madrid, 28049, Spain}
\emailAdd{sergio.ernesto.aguilar@gmail.com, a.patra@csic.es, j.pedraza@csic.es}

\abstract{We develop a new setting in the framework of braneworld holography to describe a pair of coupled and entangled uniformly accelerated universes. The model consists of two branes embedded into AdS space capping off the UV and IR regions, giving rise to a notion of dS wedge holography. Specializing in a three-dimensional bulk, we show that dS JT gravity can emerge as an effective braneworld theory, provided that fluctuations transverse to the branes are included. 
We study the holographic entanglement entropy between the branes as well as the holographic complexity within the `complexity=anything' proposal. We reproduce a Page curve with respect to an observer collecting radiation on the UV brane, as long as we take the limit where gravity decouples in that universe, thus acting as a non-gravitating bath.
The Page curve emerges due to momentum-space (UV/IR) entanglement and can be understood as analogous to the `confinement-deconfinement' transition in theories with a mass gap. Moreover, the analysis of complexity shows that the hyperfast growth phenomenon is displayed within a set of proposals, while late-time linear growth can be recovered for a different set. Our framework thus provides new test grounds for understanding quantum information concepts in dS space and dS holography.}

\maketitle

\section{Introduction}
The Anti-de Sitter/Conformal Field Theory (AdS/CFT) correspondence \cite{Maldacena:1997re,Gubser:1998bc, Witten:1998qj} provides a very fruitful test ground for developing notions of quantum gravity from our knowledge of the UV completion of spacetime. This framework is well under control when gravity in AdS is classical, which amounts to considering CFT duals with a large number of degrees of freedom, i.e., the so-called large-$N$ limit. However, we are often times interested in studying semi-classical gravitational corrections, which are of great importance in cosmology and black hole physics, among others.

Exploring the effects of $1/N$ quantum corrections within AdS/CFT involves computing loop corrections in the bulk, which are prohibitively difficult in many instances. Two alternatives to overcome this problem are: i) reducing the dimensionality of the problem, e.g., considering toy models of two-dimensional gravity where the problem of semi-classical backreaction can be solved exactly \cite{Susskind:1992gd,Russo:1992ax,Russo:1992ht,Fiola:1994ir,Fabbri:1995bz,Almheiri:2014cka,Almheiri:2019psf,Almheiri:2019qdq,Penington:2019kki,Pedraza:2021cvx,Pedraza:2021ssc}. Or ii) work in the framework of braneworld holography, where the problem of backreaction reduces to solving classical gravitational equations in one dimension higher \cite{Randall:1999ee,Randall:1999vf,Karch:2000gx,Karch:2000ct,Giddings:2000mu,Emparan:1999wa,Emparan:1999fd,deHaro:2000wj,Emparan:2002px,Rocha:2008fe,Almheiri:2019hni,Karch:2022rvr}.

\begin{figure}[t!]
\centering
    \includegraphics[width=8.3cm, trim={0 0.2cm 0 0}]{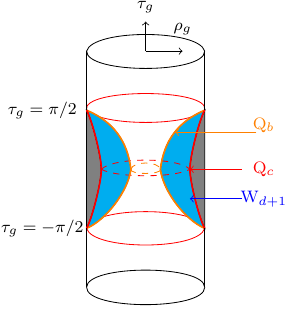}
\caption{A proposal for ``dS wedge holography'' with a pair of dS$_d$ branes, labeled by Q$_b$ (orange) and Q$_c$ (red) embedded in AdS$_{d+1}$ space with global time and radial coordinates $\tau_g$ and $\rho_g$ respectively. The branes extend between $\tau_g=\pm\tfrac{\pi}{2}$ for any brane tension, ending on a pair of codimension-2 Euclidean defects. The region between the branes (blue) is denoted as the wedge W$_{d+1}$.}
\label{fig:manybranes}
\end{figure}

In this manuscript, we construct a new setting within the framework of braneworld holography to study information-theoretic observables in de Sitter (dS) space beyond the infinite $N$ limit. Our construction uses the notion of wedge holography \cite{Akal:2020wfl,Miao:2021ual,Arias:2019zug,Miao:2020oey,Ogawa:2022fhy,Yadav:2023qfg}  originally developed for AdS Karch-Randall (KR) braneworld models \cite{Karch:2000gx,Karch:2000ct,Giddings:2000mu}. More specifically, our configuration of interest contains a pair of uniformly accelerating end-of-the-world (ETW) branes embedded in an AdS bulk spacetime, a construction that we coin ``dS wedge holography''. One of the branes is located near the asymptotic boundary, denoted the UV brane, while the other is in the interior of the bulk space, which we call the IR brane. See Fig.~\ref{fig:manybranes} for an illustration.
Similar to standard AdS wedge holography, our setup has three equivalent descriptions, and is thus \emph{doubly holographic:}
\begin{itemize}
    \item[1)] A pair of Euclidean CFT$_{d-1}$ theories on S$^1\times$S$^{d-2}$-spaces which are timelike separated from each other. We refer to these as the defect theories.
    \item[2)]  Gravity on two entangled dS$_{d}$ universes coupled to CFT$_{d}$ degrees of freedom connected at the past and future infinity via transparent boundary conditions. We refer to these as the brane theories.\footnote{Often in this paper, we will take the limit where the UV brane is placed exactly at the AdS boundary. In this case, gravity decouples from the UV theory, and it instead becomes a non-gravitating bath entangled with the IR theory.}
    \item [3)] Einstein gravity on AdS$_{d+1}$ space containing two dS$_{d}$  Randall-Sundrum branes intersecting each other only at past and future infinity.
\end{itemize}
Earlier efforts in developing holographic dictionaries for dS braneworld models with a single brane were performed in \cite{Nojiri:2002wn,Nojiri:2003jn,Calcagni:2005vn,Guo:2011qt,Emparan:2022ijy,Panella:2023lsi}. In the lower-dimensional context, models containing a pair of entangled universes have been considered in \cite{Balasubramanian:2020xqf,Balasubramanian:2020coy,Balasubramanian:2021wgd}. Meanwhile, different notions of dS wedge holography have been proposed in previous works \cite{Arias:2019zug,Miao:2020oey,Ogawa:2022fhy,Yadav:2023qfg}, although without the characteristics pursued in our work.

The particular information-theoretic quantities that we will consider are entanglement entropy and computational complexity. 
The development of the RT prescription for computing holographic entanglement entropy  \cite{Ryu:2006bv,Ryu:2006ef}, and its generalizations \cite{Hubeny:2007xt,Dong:2013qoa,Camps:2013zua,Faulkner:2013ana,Engelhardt:2014gca}, have catalyzed much of the progress in our understanding of bulk reconstruction \cite{Czech:2012bh,Balasubramanian:2013lsa,Dong:2016eik,Espindola:2018ozt,Faulkner:2018faa,Bao:2019bib} and the emergence of gravity from entanglement \cite{Lashkari:2013koa,Faulkner:2013ica,Swingle:2014uza,Faulkner:2017tkh,Haehl:2017sot,Agon:2020mvu,Agon:2021tia}, among others, giving rise to the slogan `entanglement=geometry' \cite{VanRaamsdonk:2009ar,VanRaamsdonk:2010pw,Bianchi:2012ev,Maldacena:2013xja,Balasubramanian:2014sra}. 
Entanglement, however, falls short in describing all aspects of gravitational physics, because there are particular bulk regions that entanglement surfaces cannot reach \cite{Engelhardt:2013tra}. In particular, it fails to capture the late time growth of the wormhole inside a black hole \cite{Susskind:2014moa}. A complementary effort in this direction includes the development of a holographic dual for computational complexity, a measure that diagnoses the difficulty of building a state from a reference state and a set of unitary operators (see \cite{Chapman:2021jbh} for a review). There are a number of conjectures for the holographic dual of complexity, which include the so-called ‘complexity=volume’ (CV) duality
\cite{Susskind:2014moa,Stanford:2014jda}, ‘complexity=action’ (CA) \cite{Brown:2015bva, Brown:2015lvg}, ‘complexity=volume 2.0’ (CV2.0) \cite{Couch:2016exn}, and
more recently, the ‘complexity=anything’ (CAny) dualities \cite{Belin:2021bga, Belin:2022xmt}, generalizations of the previous proposals that capture some of the basic qualitative properties
expected in complexity. Regardless of its precise dual, complexity has also shown to give rise to bulk gravitational dynamics \cite{Czech:2017ryf,Caputa:2018kdj,Susskind:2019ddc,Pedraza:2021mkh,Pedraza:2021fgp,Pedraza:2022dqi,Carrasco:2023fcj} and, thus, is expected to be able to capture aspects of bulk physics beyond entanglement.

This paper is organized as follows. In Sec.~\ref{Sec:setting} we describe the setting for our proposal of dS wedge holography. We pay particular attention to the interpretation from the point of view of the various descriptions of the system. In Sec.~\ref{sec:QI} we study holographic entanglement entropy and holographic complexity within the CV and CAny proposal for codimension-1 observables. Upon taking the limit where gravity decouples in the UV universe, we find a Page curve with respect to an observer collecting radiation in that universe, provided the IR brane is sufficiently far apart. Regarding complexity, our analysis reveals that the hyperfast growth phenomenon is displayed within a set of proposals, while late-time linear growth can be recovered for a different set. In Sec.~\ref{Sec:dS JT} we consider a refined version of our model by allowing the branes to fluctuate around their rigid locations. For a three-dimensional bulk, we find that the effective description of the branes includes dS Jackiw-Teitelboim (JT) gravity + higher order corrections that can be suppressed in certain cases. In the same Section, we extend our previous analysis of quantum information observables in the presence of brane fluctuations and we discuss their interpretation from the point of view of the effective dS JT gravity description. We conclude in Sec.~\ref{Sec:Conclusion} with a summary of our findings and some directions for future work. For the convenience of the reader, we provide a series of appendices that contain some of the technical calculations presented in the main body of the paper.

\section{UV-IR proposal for dS wedge holography}\label{Sec:setting}
In the following discussion, we will denote hatted quantities as $(d+1)$-dimensional, while un-hatted variables as $d$-dimensional. Moreover, we will adopt the following convention for the indices: $M,\,N=0,\,1,\,\dots d$, and $\mu,\,\nu=0,\,1,\,d-1$.

Our starting point is a suitable modification of a KR braneworld model \cite{Karch:2000ct,Karch:2000gx}. More specifically, we consider asymptotically AdS$_{d+1}$ Einstein gravity on a Lorentzian manifold $\hat{\mathcal{M}}$ coupled to two EOW branes following hyperbolic motion ---see Fig.~\ref{fig:manybranes}---.

The system is described by the following action
\begin{equation}\label{eq:Ihigherdim}
\begin{aligned}
    I=&\int_{\hat{\mathcal{M}}}\rmd^{d+1} x\sqrt{-\hat{g}}\qty[\tfrac{1}{2\kappa_{d+1}^2}\qty(\hat{R}-2\Lambda_{d+1})+\hat{\mathcal{L}}_{\rm bulk}]+\tfrac{1}{\kappa_{d+1}^2}\int_{\partial \hat{\mathcal{M}}}\rmd^d x\sqrt{-h}\,K\\
    &+\sum_{i}\int_{Q_i}\rmd^{d} x\,\sqrt{-h^{(i)}}\qty(\mathcal{L}^{(i)}_{\rm intrinsic}+\mathcal{L}^{(i)}_{\rm matter})\,,
    \end{aligned}
\end{equation}
where
\begin{equation}
    \Lambda_{d+1}=-\frac{d(d-1)}{2\ell_{d+1}^2}\,.
\end{equation}
In the above $Q_i$ denotes (the worldvolume) of an ETW brane; $\mathcal{L}^{(i)}_{\rm intrinsic}$ is the intrinsic theory on $Q_i$; $h^{(i)}_{\mu\nu}$ is the induced metric on $Q_i$; $\hat{\mathcal{L}}_{\rm bulk}$ is the bulk field Lagrangian density and ${\mathcal{L}}_{\rm matter}$ is the Lagrangian density of brane matter fields (\ref{eq:intr theory}). To retain analytic control over the solutions we will consider the intrinsic brane theories to be pure tensional,
\begin{equation}\label{eq:intr theory}
    \int_{Q_i}\rmd^{d} x\,\sqrt{-h^{(i)}}\mathcal{L}^{(i)}_{\rm intrinsic}=-\mathcal{T}_i\int_{Q_i}\rmd^{d} x\,\sqrt{-h^{(i)}}~,
\end{equation}
with $\mathcal{T}_i$ denoting the brane tensions, and we will set $\mathcal{L}^{(i)}_{\rm matter}=0$.\footnote{Both, changing the intrinsic brane theory or including matter fields localized on the brane would amount to a change of bulk junctions conditions, see, e.g., \cite{Dvali:2000hr,Kanda:2023zse}.} The bulk geometry will thus contain two end-of-the-world branes: one located at some arbitrary location inside the bulk, which we call the infrared (IR) brane $Q_b$, and another one closer to the asymptotic boundary, which we denote the ultraviolet (UV) brane $Q_c$.\footnote{The nomenclature alludes to energy scales of the dual theory.}

To make this construction precise, we foliate the bulk manifold with dS$_d$ slices. When bulk spacetime is given by a pure AdS$_{d+1}$, this foliation has the general form
\begin{equation}
    \rmd s_{d+1}^2=\rmd\rho^2+\sinh^2\rho\,\rmd s_{\text{dS}}^2~.\label{eq:main metric}
\end{equation}
where $\rmd s^2_{\rm dS}$ is a $d$-dimensional line element for dS space in any coordinate system.

The ``wedge" geometry $W_{d+1}$ is defined by the following limited region in the $\rho$ coordinate:\footnote{The fact that this setup can be seen as a ``wedge" in global coordinates follows from the coordinate transformation (\ref{eq:map global to dS foliation}). The branes are located at different constant-$\rho$ surfaces. Meanwhile, $\tau\rightarrow\pm\infty$ maps to the global AdS time $\tau_g=\pm\pi/2$, which corresponds to $\rho_g\rightarrow\infty$ for both branes.}
\begin{equation}
    \rho_{b}\leq\rho\leq\rho_{c}\,,
\end{equation}
where $\rho_{b}$ and $\rho_{c}$ are the locations of the ETW branes $Q_b$ and $Q_c$, respectively. Typically, the UV brane $Q_c$ will be placed very close to the asymptotic AdS boundary, such that the induced gravity on it is weak and approximately local. This brane will require a positive tension $\mathcal{T}_c>0$. Meanwhile, the IR  brane $Q_b$ is placed somewhere in the interior of the bulk geometry, such that the induced gravity on the brane is non-local, similar to the original construction by \cite{Randall:1999ee,Randall:1999vf}. As we will see below, we will require $\mathcal{T}_b<0$ to support a positive curvature on $Q_b$. Importantly, the IR and UV regions, $\rho<\rho_{b}$ and $\rho>\rho_{c}$, are integrated out using a bulk version of Wilsonian integration \cite{Heemskerk:2010hk,Faulkner:2010jy,Guijosa:2022jdo}, so that $\partial W_{d+1}={\rm Q}_b\cup {\rm Q}_c$. Such a process gives rise to induced gravity theories on the branes that take the form of Einstein gravity plus a tower of higher derivative corrections that are suppressed as the branes approach the boundary, coupled to a QFT with a large number of degrees of freedom \cite{deHaro:2000vlm,Balasubramanian:1999re,Emparan:1999pm}. The explicit form of such effective theories will not be important for the time being.

In order for the construction to hold, we require the branes to obey the Israel junction conditions \cite{Israel:1966rt} which translate to the following equations on the branes,
\begin{equation}\label{eq:israel}
K^{(i)}_{\mu\nu}-h^{(i)}_{\mu\nu} K^{(i)}=-\kappa^2_{d+1}\mathcal{T}^{(i)}\,h^{(i)}_{\mu\nu}
\end{equation}
where the indices $\mu,\,\nu$ exclude the coordinate $\rho$. Consider an outward-directed normal vector $n^\mu$, at the location of either of the branes. One can find
\begin{equation}\label{eq: extrinsic curvature Tak}
    K^{(i)}_{\mu\nu}=\tfrac{1}{2}[\partial_{n} h^{(i)}_{\mu\nu}]_{\rho^{(i)}}=\pm\coth\rho^{(i)}\,h^{(i)}_{\mu\nu}
\end{equation}
where $\partial_n=\tfrac{1}{\ell_{d+1}}\partial_\rho$. The relative $\pm$ sign indicates that $n^\mu$ has opposite directions one brane from the other. In these conventions, the positive sign is for $Q_c$, and the negative is for $Q_b$. This leads to the constraint on the brane tension
\begin{equation}\label{eq:tension}
\mathcal{T}^{(i)}=\pm\tfrac{(d-1)}{\kappa^2_{d+1}}\coth\rho^{(i)}\,,
\end{equation}
implying that $\rho^{(i)}$, $\mathcal{T}^{(i)}$ and $\ell_{d+1}$ are not all independent parameters. Notice that we do not need to add further counterterms to implement holographic renormalization because we have already integrated out the UV. These counterterms are partly responsible for giving rise to the effective theory of the UV brane. Meanwhile, the IR brane, which is by definition deep in the bulk, will be described by a non-local effective gravitational theory that arises from integrating out IR degrees of freedom. 

A few additional comments are in order. First, about the interpretation. From the branes perspective, our setting describes two \emph{entangled and coupled} universes. Entanglement generates spatial connectivity between the brane theories, which is evident from the fact that there exists a bulk spacetime region between the branes that can support non-trivial spatial correlations between the two theories. The situation is analogous to a two-sided black hole, which is understood as the dual to a thermofield double state of two entangled copies of the same theory \cite{Maldacena:2001kr}. However, the difference is that in our setup the two theories are explicitly coupled. This happens because, before integrating out, the Hamiltonian of the theory generically couples UV and IR degrees of freedom (as long as the theory is not free). The bulk manifestation of such coupling is given by the fact that one can send signals (e.g., light rays) between one brane to another, making the space in between the branes \emph{traversable}.\footnote{It would be interesting to study holographic thermodynamics of our wedge system in some detail, following the methods developed in \cite{Frassino:2022zaz}. Given the explicit coupling between the two universes, we expect a term in the first law parametrizing the energy trade-off between the two subsystems.}

Second, notice that we can alternatively glue together two exact copies of the same system along the infrared branes to form a double-sided wedge configuration, corresponding to a $\mathbb{Z}_2$-symmetric version of our setup. This process would simply amount to the change $K^{(b)}_{\mu\nu}\leftrightarrow\Delta K^{(b)}_{\mu\nu}$ in (\ref{eq:israel}), where $\Delta K_{\mu\nu}=K_{\mu\nu}^+-K_{\mu\nu}^-=2K_{\mu\nu}$,
given that $Q_b$ would be a two-sided brane, instead of a ETW brane. The corresponding brane tension would be twice the one indicated in equation (\ref{eq:tension}).

\begin{figure}[t!]
\centering
    \includegraphics[width=9cm, trim={0 0.5cm -1.2cm 0}]{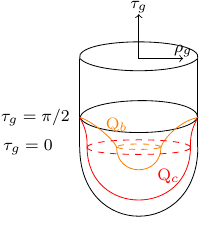}
\caption{Tunneling instanton describing a membrane creation process giving rise to our bulk state at $\tau_g=0$. The solution is analytically continued along this surface and evolved in Lorentzian signature from this point onwards.}
\label{fig:manybranes2}
\end{figure}

Lastly, a word on state preparation. In the Lorentzian setting, one should be able to prepare the bulk state by turning on sources of some (non-local) operators inserted at $\tau_g=-\tfrac{\pi}{2}$. The operators source the two branes which contract and eventually expand, hitting the boundary a second time at $\tau_g=\tfrac{\pi}{2}$, as displayed in 
Fig.~\ref{fig:manybranes}. In a Euclidean setting, however, it is evident that the branes do not reach the boundary ---see Fig.~\ref{fig:manybranes2}--- implying that we cannot prepare the state using the more standard Euclidean path integral with sources turned on, even though our setup enjoys time-reflection symmetry in Lorentzian signature. In this case, however, we can interpret the brane configuration as a tunneling instanton describing membrane creation \cite{Garriga:1993fh,Arcos:2022icf}, a higher dimensional generalization of the Schwinger effect \cite{Frob:2014zka,Fischler:2014ama}. This can be induced by turning on an antisymmetric tensor field in the boundary theory, with field strength above a certain critical value (to render the vacuum unstable).

\section{Quantum information on the dS wedge}\label{sec:QI}

\subsection{Holographic entanglement entropy}\label{Sec:Holo EE}
Let us consider an observer collecting Hawking radiation in $Q_c$. In order for this calculation to be non-trivial, we will need to take the limit where $Q_c$ approaches the asymptotic boundary, so that the bath is non-gravitating.\footnote{The key difference between gravitating and non-gravitating baths is the choice of boundary conditions for RT surfaces. It is only when the bath is non-gravitating that we can use Dirichlet boundary conditions and, hence, specify the region where we collect the radiation \cite{Geng:2020fxl}. See \cite{Ghosh:2021axl} for a discussion on the possibility of extending the Dirichlet condition to weakly gravitating baths.}
In this limit, gravity decouples and the effective theory of the UV brane is thus described by a dS QFT with a mass gap, analogous to the hard wall models of holographic QCD, e.g., \cite{Erlich:2005qh,DaRold:2005mxj}.\footnote{The entanglement entropy for a holographic dS CFT (without an infrared brane $Q_b$) was originally studied in \cite{Maldacena:2012xp,Fischler:2013fba}.} The total Hilbert space thus factorizes as\footnote{More specifically, here we use the Fock space decomposition, $\mathcal{H}=\otimes_{\vec{p}}\mathcal{H}_{\vec{p}}$, where $\mathcal{H}_{\vec{p}}$ is the Hilbert space of modes of momentum $\vec{p}$. Via holographic Wilsonian integration \cite{Heemskerk:2010hk,Faulkner:2010jy,Guijosa:2022jdo}, $\rho_b$ maps to a momentum cutoff $\Lambda$, so that we can identify $\mathcal{H}_{\text{UV}}$ and $\mathcal{H}_{\text{IR}}$ as the Hilbert spaces of modes of momentum $|\vec{p}|>\Lambda$ and $|\vec{p}|<\Lambda$, respectively. 
While this decomposition is certainly motivated 
by considering the case of free field theory, 
it applies equally well once interactions are
turned on \cite{Balasubramanian:2011wt}.}
\begin{equation}\label{totalHilbert}
    \mathcal{H}=\mathcal{H}_{\text{UV}}\otimes\mathcal{H}_{\text{IR}}\,,
\end{equation}
where $\mathcal{H}_{\text{UV}}$ is the Hilbert space of the dS QFT (with a gap) and $\mathcal{H}_{\text{IR}}$ represents the Hilbert space of the IR degrees of freedom, geometrized by $Q_b$. This IR sector contains gravity plus field theoretical degrees of freedom. 

Let $\mathbf{R}$ describe the subregion accessible to the UV observer, such that $\mathbf{R}\subset\partial\hat{\mathcal{M}}$. For simplicity, we will consider an entangling region with disk topology partitioning the $S^{d-1}$ internal space. In global dS coordinates, where
\be
\rmd s_{\text{dS}}^2=-\rmd\tau^2+\cosh^2\tau(\rmd \alpha^2+\cos^2\alpha\, \rmd \Omega_{d-2})\,,
\ee
we thus pick a constant-$\tau$ surface $\mathbf{R}=I\times S^{d-2}$, with $I=\{\alpha\in[\alpha_1,\alpha_2]\}$. Moreover, given the invariance under rotations in the $S^{d-1}$, we can fix the endpoints of $I$ in terms of a single parameter, such that $\alpha_1=-\Delta\alpha/2$ and $\alpha_2=\Delta\alpha/2$. The UV Hilbert space thus factorizes as
\begin{equation}
    \mathcal{H}_{\text{UV}}=\mathcal{H}_\mathbf{R}\otimes\mathcal{H}_{\mathbf{R}^c}~.
\end{equation}
We are interested in computing the entanglement entropy of region $\mathbf{R}$,
\be
S_\mathbf{R}=-\Tr\, (\rho_\mathbf{R}\log\rho_\mathbf{R})\,,
\ee
where $\rho_\mathbf{R}$ is the associated reduced density matrix. Holographically, we are instructed to look for codimension-2 extremal area surfaces $\mathcal{S}$ homologous to $\mathbf{R}$, such that \cite{Ryu:2006bv,Ryu:2006ef,Hubeny:2007xt}:
\begin{equation}\label{eq:Area functional}
    S_\mathbf{R}=\frac{1}{4G}\text{min}\,\underset{\mathcal{S}\sim \mathbf{R}}{\text{ext}}\mathcal{A}[\mathcal{S}]\,,\qquad \mathcal{A}[\mathcal{S}]\equiv\int_{\mathcal{S}}\rmd^{d-1}x\sqrt{\gamma}~,
\end{equation}
where $\mathcal{A}$ is the area functional and $\gamma_{\mu\nu}$ is the induced metric on the surface $\mathcal{S}$.

\begin{figure}[t!]
    \centering
\includegraphics[width=5cm,trim={0 0.8cm 0 1cm}]{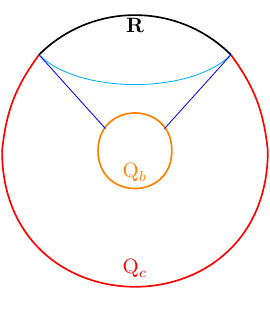}
\hspace{1.5cm}
\includegraphics[width=5cm,trim={0 0.8cm 0 1cm}]{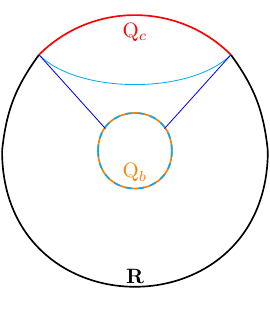}
    \caption{Constant global time slice of the dS wedge $W_{d+1}$. The black line corresponds to an entangling region $\mathbf{R}$ on Q$_c$ (in red) with an angular separation $\Delta \alpha$; while the orange circle represents Q$_b$. The cyan line represents the RT surface of the \emph{disconnected phase}, while the blue line denotes the RT surface of the \emph{connected, or island phase}. As a result of imposing a Neumann boundary condition on the IR brane, such surface will end up not lying on the same time slice, but we omit this time dependence here for simplicity. 
    In the left panel, we show the configuration for an entangling region with $\Delta \alpha<\Delta \alpha_{\text{crit}}$. In the right panel, we show the case for a region with $\Delta \alpha>\Delta \alpha_{\text{crit}}$. In this case, the homology condition forces us to add a disconnected piece to the disconnected solution that wraps the smallest section of the IR brane (at $\tau=0$).}
    \label{fig:RTs}
\end{figure}

As we increase the size of the boundary region to collect radiation, either by changing $\Delta \alpha$ or by looking at the subsystem at a later time $\tau$, the UV observer has more access to the interior of the bulk geometry. Eventually, it might seem that the observer acquires more Hawking modes than the total number of degrees of freedom on the IR brane and $\mathbf{R}^c$ jointly. Such violation in the unitary evolution might be considered a version of the information paradox for our configuration in wedge holography. To resolve the apparent paradox, one must carefully study the different phases involving RT surfaces describing the entanglement evolution, which are shown in Fig.~\ref{fig:RTs}. For small regions or early times, we expect the relevant solution to correspond to a ``disconnected'' RT surface. This solution implies there is no entanglement between $\mathbf{R}$ and the IR degrees of freedom represented by the IR ETW brane. However, as the size is increased or time evolves, we expect the appearance of a ``connected'' RT surface joining the region $\mathbf{R}$ with the IR ETW brane. The fact that these RT surfaces connect the two branes implies that there is a dominating contribution to $S_{\mathbf{R}}$ coming from UV/IR or momentum space entanglement in the original boundary theory, e.g., \cite{Balasubramanian:2011wt}. These can be interpreted as the ``island phase'' of our configuration.\footnote{Further studies on entanglement islands in cosmological settings include \cite{Krishnan:2020fer,Hartman:2020khs,Geng:2021wcq,Choudhury:2020hil,Aalsma:2021bit,Aguilar-Gutierrez:2021bns,Espindola:2022fqb,Aalsma:2022swk}.} Incidentally, we point out that this phase transition corresponds precisely to the `confinement-deconfinement' transition of the UV dS QFT \cite{Klebanov:2007ws},\footnote{More precisely, this transition was later understood to diagnose the mass gap, rather than a true confinement-deconfinement transition \cite{Jokela:2020wgs}.} discovered about two decades before than the so-called entanglement islands were discovered. We believe we are the first to draw a connection between these two phenomena. 

Finally, if we keep increasing the size of the region $\mathbf{R}$ to eventually cover more than half of the $S^{d-1}$, i.e. if we let $\Delta\alpha>\pi$, we expect the homology condition to start playing a role.
More specifically, we expect that starting from some critical angle $\Delta \alpha\geq\Delta \alpha_\text{crit}>\pi$ the relevant RT surface would go around the IR brane. To satisfy the homology constraint we must then include a disconnected piece to the RT surface that wraps the smallest section of $Q_b$ (i.e. the section at $\tau=0$). This is required to respect the Araki-Lieb inequality. As a consistency check, note that in the limit when $\mathbf{R}$ covers the full $S^{d-1}$ we should recover
\be
S_\mathbf{R}=S_{\text{UV}}=S_{\text{IR}}=S^{\text{(GH)}}_{\text{dS}}\,,
\ee
where $S^{\text{(GH)}}_{\text{dS}}$ is the Gibbons–Hawking entropy of the IR theory (which includes gravity) and we have used the fact that the overall state in the total Hilbert space (\ref{totalHilbert}) is pure. Indeed, the area of the smallest section of $Q_b$ gives precisely $S^{\text{(GH)}}_{\text{dS}}$. A similar version of this transition, known as the entanglement plateaux, occurs when the IR brane is replaced by a bulk black hole. See \cite{Hubeny:2013gta} for a detailed account of this transition.

In the following, we consider the case $\Delta\alpha \leq \pi$ for simplicity, to avoid complications with the homology condition.\footnote{In addition, note that a single observer in dS space cannot have access to the full $S^{d-1}$. Considering $\Delta\alpha > \pi$ would require the existence of a `superobserver'.} We will also focus on the $d=2$ case to make the problem analytically tractable, though, in section \ref{Sec:Conclusion} we will comment on our expectations in higher dimensional cases.

\subsubsection{Disconnected phase}
Let us first consider the disconnected solution. Since the metric of $W_{d+1}$ is just a portion of pure AdS$_{d+1}$, it is clear that the surface will lie on a constant global time in AdS. It is then convenient to work in such a coordinate system. In $d=2$, the metric of global AdS  (\ref{eq:global coordinates}) at any fixed global time $\tau_g$ is,
\begin{eqnarray}
 \rmd s^2= \rmd \rho_g^2+\sinh^2{\rho_g} \,\rmd {{\alpha}}^2~.
 \label{mbpt}
\end{eqnarray}
The RT surface in this number of dimensions is thus given by the minimal \emph{geodesic} in the $(\rho_g,\alpha)$ plane. To find it, we extremize the length functional,
\begin{equation}\label{eq:length}
    \mathcal{L}=\int\sqrt{\rho_g'^2+\sinh^2{\rho_g}}\rmd\alpha~,
\end{equation}
subject to the boundary condition
   \begin{equation}
    \rho_g(\pm \tfrac{\Delta \alpha}{2})=\rho_{g_c}\,,    
    \end{equation}
where $\rho_{g_c}\to\infty$ is the UV cutoff. Since the functional (\ref{eq:length}) does not depend explicitly on $\alpha$, then the `Hamiltonian' 
\be
\mathcal{H}=\frac{\partial\mathcal{L}}{\partial \rho_g'}\rho_g'-\mathcal{L}\,,
\ee
give us a conserved quantity, $\mathcal{H}=-E$. This allows us to explicitly solve for $\rho_g(\alpha)$. After some straightforward algebra, we find that $E=\cot(\frac{\Delta \alpha}{2})$, which yields the solution
\begin{equation}\label{eq:alpha rhog}
    \rho_g(\alpha)=\text{arcsinh}\left[\frac{\cos(\frac{\Delta \alpha}{2})}{\sqrt{\sin^2(\frac{\Delta \alpha}{2})\cos^2\alpha-\cos^2(\frac{\Delta \alpha}{2})\sin^2\alpha}}\right]\,.
\end{equation}
Next, we plug (\ref{eq:alpha rhog}) into (\ref{eq:length}) and evaluate the entanglement functional (\ref{eq:Area functional}),
\begin{equation}\label{eq:Lag d=3}
    S^{(\rm disc)}_{\mathbf{R}}=\frac{1}{4G_3}\int_{-\frac{\Delta\alpha}{2}+\delta}^{\frac{\Delta\alpha}{2}-\delta}\frac{\sin(\Delta\alpha)}{\cos(2\alpha)-\cos(\Delta\alpha)}\rmd\alpha\,.
\end{equation}
To regulate the UV divergences, the integral is cut off a $\delta$ away from the boundaries. The relation between this parameter and $\rho_{g_c}$ can be found from the solution (\ref{eq:alpha rhog}) by requiring that $\rho_g(\frac{\Delta\alpha}{2}-\delta)=\rho_{g_c}$. In the limit $\alpha\to0$, $\rho_{g_c}\to \infty$, we find
\be
\delta = 2 e^{-2 \rho_{g_c}} \cot (\tfrac{\Delta \alpha }{2})\,.
\ee
This yields
\be
S^{(\rm disc)}_{\mathbf{R}}=\frac{\rho_{g_c}+\log \left[\sin (\frac{\Delta \alpha }{2})\right]}{2 G_3}\,.
\ee
Finally, employing the map between global AdS and AdS with a global dS foliation (\ref{eq:map global to dS foliation}), we find $\rho_{g_c}=\rho_{c}+\log\, (\cosh\tau)$. Therefore
\be\label{eq:S disc total}
S^{(\rm disc)}_{\mathbf{R}}(\tau)=\frac{\rho_{c}+\log \left[\cosh\tau\sin (\frac{\Delta \alpha }{2})\right]}{2 G_3}\,.
\ee
As anticipated, the disconnected solution leads to an entanglement entropy that grows with $\tau$ without an upper bound. This leads to a non-unitary entanglement evolution according to an observer in the UV brane.

\subsubsection{Island phase}\label{Sec: Island phase}
To restore unitarity, we consider RT surfaces that connect $\mathbf{R}$ with the IR ETW brane $Q_b$. Given the rotational invariance in $d=2$ ($\alpha\to\alpha+c$), the recipe now instructs us to look for constant-$\alpha$ geodesics, possibly time-dependent. In this case, we work directly in AdS coordinates adapted to the dS foliation, which for constant-$\alpha$ reads
\be
\rmd s^2= \rmd \rho^2-\sinh^2{\rho} \,\rmd \tau^2\,.
\ee
The length functional we need to extremize is then
\begin{equation}\label{eq:lengthcon}
    \mathcal{L}=2\int\sqrt{1+\sinh^2{\rho}\,\tau'(\rho)^2}\,\rmd\rho~,
\end{equation}
subject to Dirichlet boundary conditions in the UV brane  $\tau(\rho_c)=\tau_{\text{UV}}$ and Neumann boundary conditions in the IR brane $\tau'(\rho_b)=0$. The factor of 2 in front accounts for the two geodesics, ending on the two endpoints of the interval, respectively, as shown in Fig.~\ref{fig:RTs}. The general solution for the geodesic is given by
\be
\tau(\rho)=\pm \text{arctanh}\left[\frac{\cosh \rho}{\sqrt{1+\coth^2(\tau_{\text{UV}}-c)\sinh^2\rho}}\right]+c\,.
\ee
It depends on a constant of integration, $c$. However, the IR Neumann boundary condition naturally picks $c=\tau_{\text{UV}}$ so that the solution becomes
\be\label{consttau}
\tau(\rho)=\tau_{\text{UV}}\equiv\tau\,.
\ee
Note that constant dS time $\tau$ is not equivalent to global AdS time $\tau_g$. This can be seen from the coordinate transformations (\ref{eq:map global to dS foliation}). Plugging (\ref{consttau}) into (\ref{eq:lengthcon}) we can now evaluate the entanglement functional (\ref{eq:Area functional}). The corresponding entanglement entropy in this phase yields
\begin{equation}\label{eq:Sbrane}
    S^{(\text{conn})}_{\mathbf{R}}=\frac{\rho_c-\rho_b}{2 G_3}\,.
\end{equation}
In this phase, the entanglement entropy is constant, independent of $\tau$ and $\Delta\alpha$. Since $S^{(\rm disc)}_{\mathbf{R}}$ grows without bounds at late times, clearly the connected phase will dominate at late times.

\subsubsection{Page transition}

When there are multiple RT solutions, the entanglement entropy of the region is given by the minimum between the multiple solutions. In this case, we have
\be
S_{\mathbf{R}}(\tau)=\text{min}\,\{S^{(\text{disc})}_{\mathbf{R}}(\tau),S^{(\text{conn})}_{\mathbf{R}}\}\,.
\ee
We can now compare the two phases, and determine in which cases we have a transition. To start with, notice that the disconnected phase in (\ref{eq:S disc total}) has a minimum at $\tau=0$. This is natural, since in Lorentzian signature our brane configuration exhibits time reflection symmetry. For the purposes of this section, we will thus focus on the evolution for $\tau>0$. We can imagine, for example, a tunneling event that prepares this configuration at $\tau=0$ (as explained at the end of section \ref{Sec:setting}) and then evolving the system from this time onwards.

We have already determined that at late times the connected configuration dominates, as it has less area for sufficiently large $\tau$. To guarantee a transition we need to check that the disconnected configuration dominates at early times $S^{(\text{disc})}_{\mathbf{R}}(\tau=0)<S^{(\text{conn})}_{\mathbf{R}}$. This condition implies that, in order to have a transition, we must require
\begin{equation}
\rho_b<\log\left[\csc(\tfrac{\Delta\alpha}{2})\right]\,.
\label{eq:cond4transition}
\end{equation}
Intuitively, this condition tells us that, for a fixed $\Delta \alpha$, we require a level of non-locality in the infrared dS theory (which is increased as $\rho_b$ is decreased). Only above such level of non-locality, the observer collecting radiation in $\mathbf{R}$ can then detect a Page transition. 

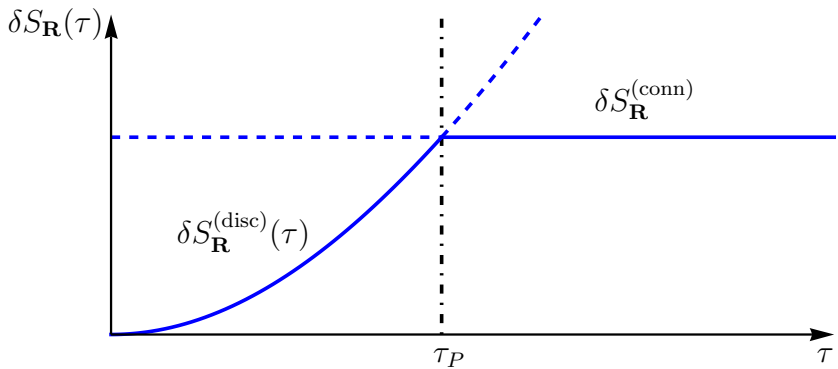
\begin{figure}[t!]
    \centering
    \includegraphics[width=0.7\textwidth,trim={0 0.5cm 0 0}]{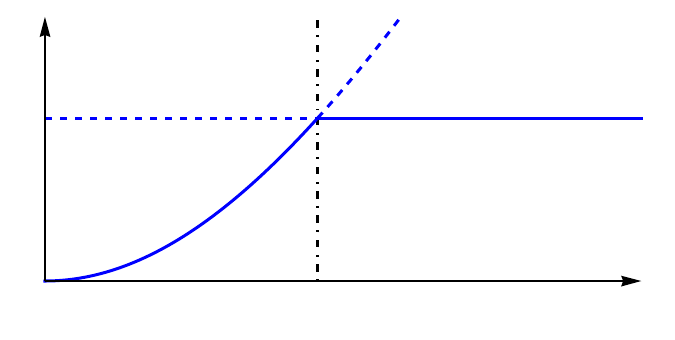}
    \begin{picture}(0,0)
\put(-332,128){$\delta S_{\mathbf{R}}(\tau)$}
\put(-268,48){$\delta S^{(\text{disc})}_{\mathbf{R}}(\tau)$}
\put(-110,98){$\delta S^{(\text{conn})}_{\mathbf{R}}$}
\put(-26,3){$\tau$}
\put(-171,3){$\tau_P$}
\end{picture}
    \caption{Evolution of entanglement entropy $\delta S_{\mathbf{R}}(\tau)=S_{\mathbf{R}}(\tau)-S_{\mathbf{R}}(0)$ for a generic boundary subregion with $\Delta\alpha<\pi$. Due to the entanglement between $\mathbf{R}$ and the IR brane degrees of freedom, the entanglement entropy saturates after Page time (\ref{pagetime}).
    \label{fig:my_label}}
\end{figure}

One can straightforwardly include boundary time dependence in the previous analysis. A quick calculation shows that, if (\ref{eq:cond4transition}) is satisfied, the Page time is given by
\be\label{pagetime}
\tau_{P}=\text{arcsech}\left[e^{\rho_b} \sin (\tfrac{\Delta \alpha }{2})\right].
\ee
The full Page curve is depicted in Fig.~\ref{fig:my_label}, for some sample parameters satisfying (\ref{eq:cond4transition}).

One might be worried that the disconnected solution ceases to exist before the Page transition takes place. This might happen because in global coordinates the maximum depth of the disconnected solution is completely fixed by $\Delta\alpha$, while the IR brane grows in time as depicted in Fig.~\ref{fig:manybranes}. Thus it is clear that after a time $\tau_*$, a disconnected solution for any given $\Delta\alpha$ will not be entirely contained within the wedge $W_{3}$. In global AdS coordinates (\ref{eq:alpha rhog}) we have that the geodesic reaches a minimum at
\be
\rho^{*}_g=\rho_g(\alpha=0)=\text{arcsinh}\left[\cot(\tfrac{\Delta \alpha }{2})\right]\,.
\ee
Translating $\rho^{*}_g$ into dS coordinates with the help of (\ref{eq:map global to dS foliation}), and equating it to $\rho_b$, we find that the RT surface intersects the brane at
\be
\tau_*=\text{arccosh}\left[\text{csch}\,\rho_b\cot (\tfrac{\Delta \alpha }{2}) \right]\,.
\ee
Fortunately, we can easily verify that $\tau_*>\tau_P$. Hence, the Page transition always takes place before the disconnected RT solution intersects the IR brane.

\subsection{Holographic complexity}\label{Sec:complexity}
There are several types of holographic complexity proposals that can be applied to our wedge holographic model \cite{Susskind:2014moa,Stanford:2014jda,Brown:2015bva, Brown:2015lvg,Couch:2016exn,Belin:2021bga,Belin:2022xmt}. For concreteness, we will focus on a set of codimension-1 `complexity=anything' proposals, first introduced in \cite{Belin:2021bga, Belin:2022xmt}. However, we will begin by illustrating the general principles with one of the simplest examples, known as the `complexity=volume' (CV) duality \cite{Susskind:2014moa,Stanford:2014jda}.

\subsubsection{CV proposal}\label{Sec:CV rigid}
CV duality states that the complexity of a quantum state is holographically dual to the maximal volume of a bulk codimension-1 hypersurface $\mathcal B$ anchored at some boundary Cauchy slice $\Sigma$ \cite{Susskind:2014moa,Stanford:2014jda},
\begin{equation}
\mathcal{C}_{\text{V}}(\Sigma)=\frac{1}{G\ell}\underset{\Sigma=\partial
\mathcal B}{\text{max}}\mathcal{V}[\mathcal{B}]\,,\qquad \mathcal{V}[\mathcal{B}]\equiv\int_{\mathcal{B}}\rmd^{d} x\,\sqrt{\gamma}~,
\label{com=vol}
\end{equation}
where $\ell$ is some arbitrary length scale, $\mathcal{V}$ is the volume functional and $\gamma_{\mu\nu}$ is the induced metric on the surface $\mathcal B$. For our purposes, we will consider a state in the full Hilbert space $\mathcal{H}_{\text{UV}}\otimes\mathcal{H}_{\text{IR}}$, so in the following we will consider surfaces that stretch between the UV and IR branes. We will discuss the boundary conditions a little below.

We recall that AdS$_{d+1}$ space foliated with dS$_{d}$ slices take the general form (\ref{eq:main metric}). Here we are interested in computing complexity in the context of static patch dS holography \cite{Susskind:2021esx,Chapman:2021eyy,Jorstad:2022mls,Aguilar-Gutierrez:2023zqm}, thus, we will start with the dS metric in the static patch.\footnote{In the foliation (\ref{eq:main metric}), static patches of dS$_d$ correspond to Rindler wedges of AdS$_{d+1}$.} In this case the bulk metric takes the form (\ref{eq:main metric}) with
\begin{equation}
    \rmd s^2_{\text{dS}}=-f(r)\,\rmd t^2+\tfrac{\rmd r^2}{f(r)}+r^2\rmd\Omega_{d-2}^2\,,\qquad f(r)=1-H^2r^2\,,
\end{equation}
where $H$ is the Hubble constant.\footnote{From here on we will set $H= 1$. We can easily restore $H$ anywhere in our calculations by simple dimensional analysis.} The idea of static patch holography is to consider two antipodal static patches in dS (Left/Right), which are maximally entangled. The situation is very similar to a two-sided AdS black hole \cite{Maldacena:2001kr}, because the space in between the two patches is understood as emerging purely from entanglement \cite{Susskind:2021esx}. We can imagine having a pair of such static patches in the UV and IR brane theories, and in fact, at each constant-$\rho$ bulk slice. To have a notion of time evolution, we regulate the null horizons by introducing timelike \emph{stretched horizons}. The surfaces $\mathcal{B}$ can then be anchored along the stretched horizons at given times $t_R$ and $t_L$, and  $\rho\in[\rho_b,\rho_c]$. 

\begin{figure}
    \centering
    \includegraphics[width=0.7\textwidth]{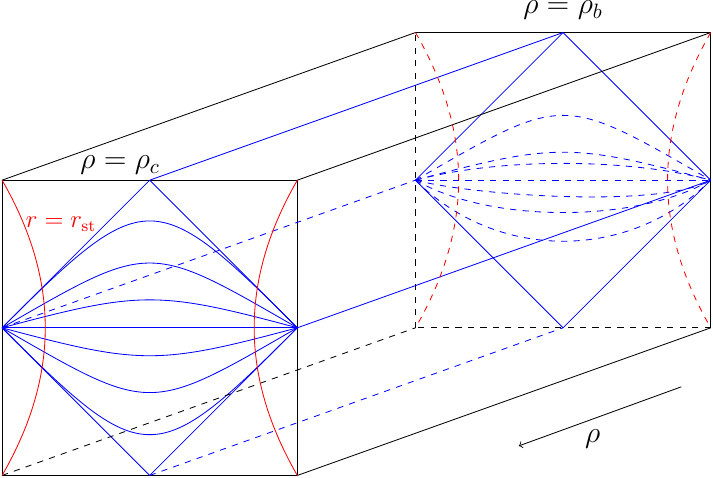}
    \caption{Representation of maximal volume surfaces $\mathcal{B}$, stretching between the stretched horizons (shown in red) and the UV ($\rho=\rho_c$) and IR ($\rho=\rho_b$) branes.}
    \label{fig:Corrections to Extremal surfaces}
\end{figure}

We are mostly interested in symmetric configurations with 
\begin{equation}\label{eq:time evol}
    t_R=t_L\equiv\frac{t}{2}~.
\end{equation}
Since the relevant surfaces go through an inflating region complementary to the two static
patches, it is convenient to introduce Eddington-Finkelstein (EF) coordinates ($v$, $r$) by the change of variable $\rmd t=\rmd v-\rmd r/f(r)$,
\be\label{eq:EF metric}
\rmd s^2_{\text{dS}}= -f (r)\rmd v^2 + 2\rmd v\rmd r + r^2\rmd\Omega_{d-2}^2\,.
\ee 
We consider general profiles for the maximal volume surfaces, parametrized by functions $r(\sigma,\,\rho)$, $v(\sigma,\,\rho)$. The volume functional is thus given by:
\begin{equation}\label{eq:modified vol Funct}
    \mathcal{V}=\Omega_{d-2}\int\rmd\rho\,\rmd\sigma\,r^{d-2}\sinh^{d-1}\rho\,\sqrt{\Delta}\,,
\end{equation}
where
\begin{equation}\label{eq:Delta}
    \begin{aligned}
        \Delta\equiv&\,\,\partial _{\sigma } v(\sigma ,\rho ) (-\partial _{\sigma } v(\sigma ,\rho ) f(r(\sigma ,\rho ))+2 \partial _{\sigma } r(\sigma ,\rho ))\\
        &-\sinh ^2(\rho ) (\partial _{\sigma } r(\sigma ,\rho ) \partial _{\rho }v(\sigma ,\rho )- 
   \partial _{\rho } r(\sigma ,\rho ) \partial _{\sigma }v(\sigma ,\rho )){}^2~.
    \end{aligned}
\end{equation}
From this functional, one may derive the equations of motion, and proceed to solve them numerically. The solutions determine a family of hypersurfaces that stretch between the stretched horizons and between the UV and IR branes, with some given $\rho$ profiles.   
See Fig.~\ref{fig:Corrections to Extremal surfaces} for an illustrative sketch.
Alternatively, one may proceed analytically, however, resorting to some kind of perturbative scheme. To be concrete, will proceed as follows in the remainder of this section: we will send $Q_c$ to the AdS boundary (as we did for the entanglement entropy calculation) and assume that $\rho_b$ is large but finite. In this limit, we are left with a weakly gravitating theory in the IR brane coupled to a non-gravitating UV bath. To leading order, then, we can assume that the embedding functions are $\rho$-independent, with corrections appearing in powers of $e^{-\rho}$.

\subsubsection*{Leading order contribution}
In a near boundary expansion, the zeroth order solutions are surfaces that do not depend explicitly on $\rho$. In that case, the volume functional (\ref{eq:modified vol Funct}) can be expressed as
\begin{equation}\label{eq:Vol d}
    {\mathcal{V}}=2\Omega_{d-2}I_{(d-1)}(\rho_c,\,\rho_b)\int_{r_{\rm st}}^{r_t} \rmd \sigma\,r^{d-2}\sqrt{-f(r)\,\dot{v}(\sigma)^2+2\dot{v}(\sigma)\dot{r}(\sigma)}~,
\end{equation}
where dotted quantities indicate derivatives with respect to the parameter $\sigma$, $r_t$ is the turning point of the functional, i.e., the location where $\dot{r}=0$,
\begin{equation}\label{eq:ftilde}
    I_{(d)}(\rho_c,\,\rho_b)\equiv\int^{\rho_c}_{\rho_b}\rmd\rho\,\sinh^{d}\rho~;
\end{equation}
and $r_{\rm st}$ the stretched horizon. The latter represents a preferred location within the static patch of dS space where gravitational dressing of observables (e.g., the CAny proposals) can be performed. See \cite{Jorstad:2022mls,Anegawa:2023wrk,Baiguera:2023tpt,Anegawa:2023dad,Aguilar-Gutierrez:2023zqm} for discussions on holographic complexity and stretched horizons in static patch dS holography.

Note that at the leading order, (\ref{eq:Vol d}) is in fact proportional to the standard volume in dS space, up to the factor $I_{(d-1)}(\rho_c,\,\rho_b)$ that captures the contribution to the volume transverse to the branes. The result is known in the literature, e.g., \cite{Jorstad:2022mls},
\begin{equation}
\begin{aligned}
t&=-2\int_{r_{\rm st}}^{r_t}\tfrac{\rmd r\,P_v}{f(r)\sqrt{P_v^2+f(r)r^{2(d-2)}}}~,\\
        {\mathcal{V}}&=2\Omega_{d-2}I_{(d-1)}(\rho_c,\,\rho_b)\int_{r_{\rm st}}^{r_t}\rmd r\tfrac{r^{2(d-2)}}{\sqrt{P_v^2+f(r)r^{2(d-2)}}}~,\label{eq:Vol2 integrated}
\end{aligned}
\end{equation}
where
\begin{equation}\label{eq:conserved Pv}
P_v=(r_t^2-1)r_t^{d-2}
\end{equation}
parametrizes the implicit relation ${\mathcal{V}}(t)$ in (\ref{eq:Vol2 integrated}).

Upon close inspection, it can be seen that the rate of growth of volume near the critical time $t_\infty\equiv t(P_v\rightarrow\infty)=2\,\text{arctanh}\, r_{\rm st}$ experiences a divergence, leading to the hyperfast growth phenomenon \cite{Jorstad:2022mls}. For example, for $d=2$ and to leading order in a near boundary expansion, the maximal volume and its growth around the critical time are given by
\begin{equation}\label{eq:late t}
\begin{aligned}
\lim_{P_v\rightarrow\infty}{\mathcal{V}}&=\mathcal{V}_0=\pi\qty(\cosh\rho_c-\cosh\rho_b)~,\\
\lim_{P_v\rightarrow\infty}       \dv{\mathcal{V}}{t}& =\sqrt{\tfrac{r_{\rm st}}{t_\infty-t}}\qty(\cosh\rho_c-\cosh\rho_b)+\ldots~,
\end{aligned}
\end{equation}
where the ellipsis denotes terms that remain finite in such a limit. Although the rate of growth of complexity diverges as one takes the limit $t\to t_{\infty}$ ($P_v\rightarrow\infty$), the complexity itself remains finite. We will see later in Sec.~\ref{eq:Holo complexity dilaton} that this is no longer true for dS JT gravity, in which case, complexity as well as its growth rate diverge as one approaches the critical time $t_{\infty}$.\footnote{In \cite{Jorstad:2022mls} the authors proposed to introduce a regulator surface for $t\geq t_\infty$ in order to study the evolution of this observable at later times. We will not pursue this avenue here.}

\subsubsection*{Corrections to the maximal volume surfaces}
Our focus is now to find corrections to the complexity to account for the separation of the branes $\rho\in\qty[\rho_b,\,\rho_b]$ in a near boundary expansion. To make the procedure explicit, we restrict our attention to the case of $d=2$. We use the following Ansatz for the embedding functions describing the maximal surface slice $\mathcal{B}$,
\begin{equation}\label{eq:Ansatz CV}
\begin{aligned}
    r(\rho,\,\sigma)&\simeq r_0(\sigma)+e^{-a\rho}r_1(\sigma)+\cdots~,\\
    v(\rho,\,\sigma)&\simeq v_0(\sigma)+e^{-b\rho}v_1(\sigma)+\cdots~,
\end{aligned}
\end{equation}
where $r_0(\sigma)$ and $v_0(\sigma)$ are the zeroth order solutions \cite{Jorstad:2022mls},
\begin{equation}
    \begin{aligned}
        r_0(\sigma)&=\sqrt{1+P_v^2}\sin\sigma~,\label{eq:correction r}\\
        v_0(\sigma)&=\tfrac{1}{2}\log\qty[\tfrac{(1+\sqrt{1+P_v^2}\sin\sigma)(1-P_v\tan\sigma)}{(1-\sqrt{1+P_v^2}\sin\sigma)(1+P_v\tan\sigma)}]~;
    \end{aligned}
\end{equation}
and $a$ and $b$ are constants, which can be determined from the equations of motion. Plugging such an Ansatz into the Euler-Lagrange equations resulting from the functional (\ref{eq:modified vol Funct}) and expanding for large $\rho$ we obtain a set of coupled equations for $r_1(\sigma)$ and $v_1(\sigma)$.
In general, the solutions are quite complicated, but in the hyperfast regime, i.e. $P_v\rightarrow\infty$, we find that
\begin{align}
    r(\rho,\,\sigma)=&r_0(\sigma)+e^{-3\rho}\cos\tfrac{\sigma}{2}\qty(a_re^{\sqrt{\tfrac{7}{2}}\sigma}+b_re^{-\sqrt{\tfrac{7}{2}}\sigma})+\mathcal{O}(e^{-6\rho})~,\label{eq:r(sigma rho)}\\
    v(\rho,\,\sigma)=&v_0(\sigma)+e^{-3\rho}\sec\sigma\qty(a_ve^{\sqrt{\tfrac{7}{2}}\sigma}+b_ve^{-\sqrt{\tfrac{7}{2}}\sigma})+\mathcal{O}(e^{-6\rho})~,\label{eq:v(sigma rho)}
\end{align}
where $a_r$, $b_r$, $a_v$, and $b_v$ are integration constants that can be suitably fixed from the boundary conditions. Regarding the latter, we follow \cite{Jorstad:2022mls} and pick a parametrization
such that $\sigma\in\qty[0,\,\tfrac{\pi}{2}]$ is mapped to the region $r\in\qty[0,\,r_{t}]$, where $r_t=\sqrt{1+P_v^2}$ denotes the turning point (obtained upon setting $d=2$ in (\ref{eq:conserved Pv})). Thus we impose
\begin{equation}\label{eq:conditions infty}
    \begin{aligned}
        r(\rho,\,\sigma=0)=0~,\qquad v(\rho,\,\sigma=\pi/2)&=r^*(r_t)~.
    \end{aligned}
\end{equation}
The Zeroth order solutions already satisfy these conditions. In order for (\ref{eq:conditions infty}) to hold at the next order we then need to impose 
\begin{equation}\label{eq:conditions infty1}
    \begin{aligned}
        r_1(\rho,\,\sigma=0)=0~,\qquad v_1(\rho,\,\sigma=\pi/2)&=0~.
    \end{aligned}
\end{equation}
This yields
\be
b_r=-a_r\,,\qquad b_v=a_v=0\,.
\ee
Finally, we fix the remaining constant by maximizing the volume functional at the given order, subject to the constraint that the surface must be spacelike everywhere. More specifically, we require that $\det(h_{\mu\nu})\geq 0$ for $\rho\in[\rho_b,\rho_c]$. This leads to 
\be\label{eq:modified vol ar}
a_r\leq -\sqrt{\tfrac{2}{7}}P_v e^{3\rho_b}\,,
\ee
as $P_v\to\infty$. It can be shown that the volume functional is maximized when (\ref{eq:modified vol ar}) is saturated, so the final result for the embedding yields
\begin{align}\label{eq:embedding corrected1}
    r(\rho,\,\sigma)=&r_0(\sigma)-\sqrt{\tfrac{2}{7}}P_v e^{-3(\rho-\rho_b)}\cos\tfrac{\sigma}{2}\qty(e^{\sqrt{\tfrac{7}{2}}\sigma}-e^{-\sqrt{\tfrac{7}{2}}\sigma})+\mathcal{O}(e^{-6\rho})~,\\
    v(\rho,\,\sigma)=&v_0(\sigma)+\mathcal{O}(e^{-6\rho})~.\label{eq:embedding corrected2}
\end{align}
Notice that $r_0(\sigma)\propto P_v$ in the limit $P_v\to\infty$ so it is not strange that the correction $r_1(\sigma)$ also displays the same scaling. Also, note that the factor $e^{-3(\rho-\rho_b)}=1$ in the IR so the correction seemingly competes with the leading order result there. This is completely fine. Upon close inspection, we find that the volume functional (\ref{eq:modified vol Funct}) becomes\footnote{We define $\sigma_{\text{st}}$ such that $r(\rho,\sigma_{\text{st}})=r_{\text{st}}$. At the leading order, $\sigma_{\text{st}} = \arcsin\tfrac{r_{\rm st}}{P_v}\to 0$ as $P_v\to \infty$.}
\begin{equation}\label{eq:CV corrected}
\begin{aligned}
\mathcal{V}&=\mathcal{V}_0+\tfrac{e^{-2\rho_b}}{4}\int_{\sigma_{\text{st}}}^{\frac{\pi}{2}}\rmd\sigma~ \tfrac{r_1'v_0'+r_0r_1{v_0'}^2}{\sqrt{2v_0'r_0'-f(r_0){v_0'}^2}}+\mathcal{O}(e^{-5\rho_b})\\
&\simeq\mathcal{V}_0-\mathcal{V}_1~,\qquad \mathcal{V}_1\equiv\tfrac{\rme^{\rho_b}}{2\sqrt{7}}\sinh(\sqrt{\tfrac{7}{2}}\tfrac{\pi}{2})
\end{aligned}
\end{equation}
where we have used the explicit corrected embeddings (\ref{eq:embedding corrected1})-(\ref{eq:embedding corrected2}). Note that the correction with respect to the zeroth order result (\ref{eq:late t}) is suppressed in the $P_v\rightarrow\infty$ regime since $\mathcal{V}_0$ contains divergences of order $e^{\rho_c}$, while $\mathcal{V}_1$ is finite as $\rho_c\to \infty$. Finally, since $d\mathcal{V}/dt\to\infty$ as $P_v\to\infty$ at the leading order, we expect on general grounds that this would remain the same while going to higher orders. Technically, to check the time derivative in such a limit would require us to compute corrections to the embeddings beyond $P_v\to \infty$ which falls beyond our analytical control. However, it is highly unlikely that the corrections would \emph{exactly} cancel the leading order divergence. Hence, we conclude that the hyperfast growth phenomenon is not affected by higher order corrections, other than a shift in $\mathcal{V}_0$. 

\subsubsection{`Complexity=Anything' proposal}

Next, we consider how to incorporate different types of evolution within the complexity=anything proposals, as recently done in \cite{Aguilar-Gutierrez:2023zqm}. We will focus on codimension-1 observables within the class of the CAny proposal, introduced in \cite{Belin:2021bga,Belin:2022xmt},
\begin{equation}\label{eq:Volepsilon}
    \mathcal{C}^\epsilon \equiv \frac{1}{G_N}\int_{\Sigma_\epsilon}\rmd\rho\rmd^{d-1}\sigma\,\sqrt{\gamma}~F[g_{\mu\nu},\,\mathcal{R}_{\mu\nu\rho\sigma},\,\nabla_\mu]~,
\end{equation}
where $F[g_{\mu\nu},\,\mathcal{R}_{\mu\nu\rho\sigma},\,\nabla_\mu]$ is an arbitrary scalar functional of curvature invariants in the bulk AdS$_{d+1}$ space,\footnote{The simplest proposal would be to set $F=1$, which corresponds to the volume of the $\Sigma_\epsilon$ slices.} $\Sigma_\epsilon$ is a $d$-dimensional spatial slice, and $\gamma$ is the determinant of the induced metric on $\Sigma_\epsilon$. 

\begin{figure}[t!]
    \centering
    \includegraphics[width=0.35\textwidth]{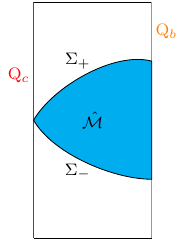}
    \caption{Region employed in the codimension-one `Complexity=Anything' proposal. The codimension-zero bulk region in AdS$_{d+1}$ space is denoted $\hat{\mathcal{M}}$, and $\Sigma_+$ and $\Sigma_+$ are the future and past boundary slices respectively, which are anchored on the ETW branes Q$_b$ and Q$_c$.}
    \label{Fig:pureAdS}
\end{figure}

Following \cite{Aguilar-Gutierrez:2023zqm}, we evaluate the codimension-one observable $\mathcal{C}^{\epsilon}$ on the future/past slices $\Sigma_{\epsilon}=\Sigma_{\pm}$ of the codimension-zero region obtained by extremizing another functional, $\mathcal{C}_{\rm CMC}$. To define such a functional, we employ a combination of co-dimension one and co-dimension zero volumes with different weights,
\begin{equation}\label{eq:regions CMC}
\begin{aligned}
    \mathcal{C}_{\rm CMC}&\equiv\tfrac{1}{G_N}\biggl[\alpha_+\int_{\Sigma_+}\rmd\rho\rmd^{d-1}\sigma\,\sqrt{\gamma}+\alpha_-\int_{\Sigma_-}\rmd\rho\rmd^{d-1}\sigma\,\sqrt{\gamma}+\alpha_B\int_{\hat{\mathcal{M}}}\rmd^{d+1}x\sqrt{-g}\biggr]\,,
\end{aligned}
\end{equation}
where $\hat{\mathcal{M}}$ is a $(d+1)$-dimensional bulk region; $\alpha_\pm$, $\alpha_B$ are constants; and $\Sigma_\pm$ are the $d$-dimensional future and past slices, such that $\partial\hat{\mathcal{M}}=\Sigma_+\cup\Sigma_-$. See Fig.~\ref{Fig:pureAdS} for an illustration.
Further, the extremization of $\mathcal{C}_{\rm CMC}$ reveals that $\Sigma_\pm$ are constant mean curvature (CMC) slices, whose mean curvature, $K$, is given by:
\begin{equation}
    K_\epsilon\equiv\eval{K}_{\Sigma_\epsilon} = -\epsilon\frac{\alpha_B}{\alpha_\epsilon}~.
\end{equation}
where we take the normal vector to be timelike future directed. 

We will now perform a similar analysis to Sec. \ref{Sec:CV rigid} to evaluate the functional (\ref{eq:Volepsilon}). That is, we will send $Q_c$ to the AdS boundary and assume that $\rho_b$ is large but finite so that the leading order embeddings are $\rho$-independent. The simplest observable corresponds to setting $F=1$ in (\ref{eq:Volepsilon}). In this case, the functional becomes
\begin{equation}
   \mathcal{C}^\epsilon=\tfrac{\Omega_{d-2}}{G_N}\sum_{\epsilon=\qty{+,-}}\int_{\Sigma_\epsilon}\rmd\rho\,\rmd\sigma\,r^{d-2}\sinh^{d-1}\rho\,\qty(\sqrt{\Delta}-\tfrac{r\sinh\rho K_{\epsilon}}{d-1}\partial_\sigma v)\,,
\end{equation}
where $\Delta$ is given in (\ref{eq:Delta}). From the previous expression, we note that the leading contributions are $\mathcal{O}(e^{d\rho_c})$, while the leading $\rho$-dependent terms will be $\mathcal{O}(e^{-(d+1)\rho_b})$. We will focus on the observables that display late-time growth at the leading order in the regime of interest. 

The details on the evaluation of the leading order late-time evolution of complexity can be found in the App. \ref{App:CAny details}. The result is: 
\begin{equation}\label{eq:Vol late times}
    \lim_{t\rightarrow\infty}\dv{t} \mathcal{C}^\epsilon \simeq I_{d-1}(\rho_c,\,\rho_b)\frac{\Omega_{d-2}}{G_N}\sqrt{-f(r_{f}) r_{f}^{2(d-2)}}+\dv{P_v^\epsilon}{t}\qty[\dots]\,,
\end{equation}
where $r_{f}\equiv\lim_{t\rightarrow \infty}r_{t}$, and the dots indicate the remaining integral in (\ref{eq:complexity growth}). The $\dv{P_v^\epsilon}{t}$ term vanishes provided that the effective potential (\ref{eq:potential arbitrary}) reaches a maximum, from which it follows that
\begin{equation}\label{eq:rt location}
   4 r_{f}f\left(r_{f}\right)\left((d-2)f'\left(r_{f}\right)+k^\epsilon_{d}(\rho_c,\,\rho_b)^2 r_{f}\right)+4(d-2)^2f\left(r_{f}\right){}^2+r_{f}^2f'\left(r_{f}\right){}^2=0~,
\end{equation}
with $k^\epsilon_d(\rho_c,\,\rho_b)$ in (\ref{eq:k^epsilon_d}). Since we are considering $f(r)=1-r^2$ then the solution to (\ref{eq:rt location}) is just:
\begin{equation}\label{eq:Key to crit}
    r_f^2=\frac{k^\epsilon_d(\rho_c,\,\rho_b)^2-2(d-1)(d-2)\pm\abs{k^\epsilon_d(\rho_c,\,\rho_b)}\sqrt{k^\epsilon_d(\rho_c,\,\rho_b)^2-4(d-2)}}{2(k^\epsilon_d(\rho_c,\,\rho_b)^2-(d-1)^2)}~,
\end{equation}
for which we require $\abs{k^\epsilon_d(\rho_c,\,\rho_b)}\geq2\sqrt{d-2}$ $\forall d>2$ for the late-time linear growth (\ref{eq:Vol late times}) to be valid. For the $d=2$ case with $r_{\rm st}\neq0$, there is a single turning point 
\begin{equation}
    r_f=\qty(1-\tfrac{1}{k^\epsilon_d(\rho_c,\,\rho_b)^2})^{-1}\,.
\end{equation}
We then require $k^\epsilon_d(\rho_c,\,\rho_b)\geq1$ for the late-time growth to be realized. In particular, when $\abs{k^\epsilon_d(\rho_c,\,\rho_b)}=1$, the growth can be exponentially increasing \cite{Aguilar-Gutierrez:2023zqm}. In contrast, the CAny codimension-1 proposals that do not satisfy the bound on $\abs{k^\epsilon_d(\rho_c,\,\rho_b)}$ will display similar behavior to what we discussed for CV, namely that there will be a hyperfast growth for complexity \cite{Aguilar-Gutierrez:2023zqm}, which is $\lim_{t\rightarrow t_\infty}\dv{C_V}{t}\rightarrow\infty$ at finite times.\footnote{Notice that these results are not sensitive to the particular location of the stretched horizon $r_{\rm st}$, and one might as well set $r_{\rm st}=0$, as for a static patch observer. For further discussions about the class of CAny observables in asymptotically dS space refer to \cite{Aguilar-Gutierrez:2023zqm}.} It would be interesting to incorporate the $\rho$-dependent corrections in the extremal surfaces, but we leave such analysis for future work.

\section{JT gravity from the fluctuating dS wedge}\label{Sec:dS JT}

\subsection{Effective action from fluctuating branes}\label{Sec:dS JT action}

We are now interested in adding dynamics to the system when studying quantum information measures that are exactly solvable in lower dimensions. There are different approaches that one can take, such as adding an intrinsic gravity term in (\ref{eq:intr theory}); or allowing the two branes to fluctuate about their rigid locations. We opt for the latter. For AdS$_2$ branes embedded in the ambient AdS$_3$ space-time, small transverse fluctuations give rise to interesting dynamics in the intermediate holographic description \cite{Geng:2022slq,Geng:2022tfc,Deng:2022yll}. Specifically, it leads to 2d JT gravity.\footnote{More general fluctuations give rise to other interesting effective 2d theories \cite{FNSS}. Alternatively,
one could add intrinsic brane dynamics via DGP-like terms to obtain JT or other brane theories \cite{Chen:2020uac}.} We will perform a similar analysis in this section for our dS setup.\footnote{Previously, dS JT gravity has been also derived in flat wedge holography in \cite{Bhattacharjee:2022pcb}.} Particularly, we are interested in determining the role of the scalar mode ---the radion--- in the effective dynamics of the IR and UV dS$_2$ branes embedded in bulk AdS$_3$ with arbitrary transverse fluctuations.

As before, we use hatted variables to describe quantities in 3d bulk and un-hatted ones for 2d quantities. The dS$_2$ foliation of AdS$_3$ can then be written as
\be
    \rmd s_{2+1}^2=\hat{g}_{MN}\rmd x^M\rmd x^N=\rmd\rho^2+\sinh^2\rho\, (g_{\mu\nu}\rmd x^\mu\rmd x^\nu) ~.
\ee
The Ricci scalar can be decomposed as follows
\begin{equation}
    \hat{R}[\hat{g}]=e^{-2\omega}R[g]-6\qty(\partial\omega)^2-4\square\omega\,.
\end{equation}
Here $e^\omega=\sinh\rho$, hence
\begin{equation}
    \hat{R}[\hat{g}]=\csch^2\rho \,R[g]-4-2\coth^2\rho\,,
\end{equation}
and $\sqrt{-\hat{g}}=\sinh^2\rho\sqrt{-g}$. The bulk action with two branes embedded reads
\begin{align}
    I=&-\frac{1}{2\kappa_3^2}\int_{\hat{\mathcal{M}}}\rmd^3x\sqrt{-\hat{g}}\qty(\hat{R}+2)-\frac{1}{\kappa_3^2}\sum_{i=b,\,c}\int_{Q_i}\rmd^2x\sqrt{-h_i}(K_i-\mathcal{T}_i)\,,
    \end{align}
    with $h_{\mu\nu}=\sinh^2\rho\,{g}_{\mu\nu}$. We now let the branes fluctuate about their rigid locations, i.e.,
\begin{equation}\label{eq:brane fluctuations}
    \rho_{b,\,c}(x^{\mu})=\rho_{b,\,c}+\delta\varphi_{b,\,c}(x^\mu)~,
\end{equation}
Using equations (\ref{eq: extrinsic curvature Tak})-(\ref{eq:tension}) we now obtain
\begin{align}\label{eq:minus sign check}
    I=&-\frac{1}{2\kappa_3^2}\int\rmd^2x\int_{\rho_b+\delta\varphi_b}^{\rho_c+\delta\varphi_c}\rmd\rho\sqrt{-g}\qty( R[g]-2\qty(\sinh^2\rho+\cosh^2\rho))\nonumber\\
    &-\frac{1}{\kappa_3^2}\int\rmd^2x\sqrt{-g}\,\sinh^2(\rho_c+\delta\varphi_c)\qty(2\coth(\rho_c+\delta\varphi_c)-\coth(\rho_c))\nonumber\\
    &+\frac{1}{\kappa_3^2}\int\rmd^2x\sqrt{-g}\,\sinh^2(\rho_b+\delta\varphi_b)\qty(2\coth(\rho_b+\delta\varphi_b)-\coth(\rho_b))~.
\end{align}
The relative sign in the last 2 integrals is due to the direction of the normal vector at the location of the UV vs. IR branes. Further simplification yields:
\begin{equation}
    I=-\frac{1}{2\kappa_3^2}\int\rmd^2x\sqrt{-g}\eval{\qty[(\rho_i+\delta\varphi_i)R[g]-2\Lambda_i]}_{i=b}^{i=c}~,
\end{equation}
where
\begin{equation}
\begin{aligned}
    \Lambda_i&=\frac{1}{2}\sinh\qty[2(\rho_i+\delta\varphi_i)]\qty[\tfrac{\coth(\rho_i)}{\coth(\rho_i+\delta\varphi_i)}-1]\\
    &=\delta\varphi_i+\coth(\rho_i)\,\delta\varphi_i^2+\mathcal{O}(\delta\varphi^3)
\end{aligned}
\end{equation}
Then, we find dS$_2$ JT gravity at the first order in perturbation theory,
\begin{equation}\label{eq:action dS JT}
    I=-\frac{\varphi_0}{2\kappa_3^2}\int\rmd^2x\sqrt{-g}R[g]-\frac{1}{2\kappa_3^2}\int\rmd^2x\sqrt{-g}\varphi\big(R[g]-2\big)~.
\end{equation}
where we have introduced the notation
\begin{equation}\label{eq:phi0}
    \varphi_0=\rho_c-\rho_b\,,\qquad \varphi=\delta\varphi_c-\delta\varphi_b\,.
\end{equation}
This is precisely the action for the pure dS JT gravity, which appears here as an effective description for the wedge system with two end-of-the-world branes.
The quadratic order term in the effective action is found to be 
\begin{equation}
\begin{aligned}
        I_{\text{quadratic}}=-\frac{1}{\kappa_3^2}\int \rmd^2 x\sqrt{-g}\Big[&\frac{\coth{\rho_c}}{2}\nabla_{\mu}\delta\varphi_c\nabla^{\mu}\delta\varphi_c-\coth{\rho_c}(\delta\varphi_c)^2\\
    &-\frac{\coth{\rho_b}}{2}\nabla_{\mu}\delta\varphi_b\nabla^{\mu}\delta\varphi_b+\coth{\rho_b}(\delta\varphi_b)^2\Big]~.
\end{aligned}
\end{equation}
Following the arguments of \cite{Geng:2022tfc,Geng:2022slq}, one can eliminate $I_{\text{quadratic}}$ in the bulk action by imposing anti-symmetric fluctuations $\delta\varphi_c=-\delta\varphi_b$ and then performing a Weyl transformation of the 2$d$ metric,\footnote{Anti-symmetric fluctuations naturally occur for a double-sided or $\mathbb{Z}_2$-symmetric wedge construction. Alternatively, the contribution from the $I_{\text{quadratic}}$ could be neglected by keeping the two branes very close to each other such that $\varphi(x)\ll\varphi_0\ll1$.}
\be
    g_{\mu\nu}\to e^{\frac{\mathcal{T}_1\mathcal{T}_2}{\mathcal{T}_1+\mathcal{T}_2}\varphi}g_{\mu\nu}\,.
\ee
If we further take the UV brane $Q_c$ to the boundary (as we did in the previous section) its fluctuations would be suppressed and we could set $\delta\varphi_c=0$ and $\delta\varphi_b\equiv-\varphi$.
In this case, the pure dS JT gravity would correspond to the effective dynamics of the IR brane $Q_b$, while the theory on $Q_c$ would simply correspond to a non-gravitating bath coupled to the JT sector.

Either way, from the action (\ref{eq:action dS JT}) we recover the sourceless equations of motion of dS JT gravity,
\begin{align}
R&=2\,,\\
0&=(\nabla_\mu\nabla_\nu-g_{\mu\nu}\nabla^2-R_{\mu\nu})\varphi\,.\label{eq:Dilaton eq}
\end{align}
The vacuum solution in global coordinates is:
\begin{equation}\label{eq:Sol to dS JT EOM}
    \begin{aligned}
        \rmd s^2&=-\rmd\tau^2+\cosh^2\tau\,\rmd\alpha^2~,\\
        \varphi(x)&=\varphi_r\,\sin\alpha\,\cosh\tau
    \end{aligned}
\end{equation}
with $\alpha\sim\alpha+2\pi$, $\tau\in\qty(-\infty,\,\infty)$, and $\varphi_r\in\mathbb{R}$ a constant.

Notice that the addition of the scalar mode breaks asymptotic isometries of dS$_2$ space, namely the reparametrization of coordinates at $\mathcal{I}^+$ \cite{Maldacena:2019cbz}.

\subsection{Quantum information on the effective theory}\label{Sec:QI JT}

Given the effective 2d dilaton gravity theory arising from fluctuating branes, we will now explore quantum information aspects in such an intermediate description.

\subsubsection{Holographic entanglement entropy}\label{Sec:Dilaton HEE}
In Sec. \ref{Sec:Holo EE} we explored the holographic entanglement entropy from the point of view of the UV theory. This calculation is non-trivial when we send $Q_c$ to the boundary, and thus the theory acts as a non-gravitating bath. In this section, we may take the same limiting case, such that the effective JT description describes the dynamics of the IR theory. However, we will be concerned with an observer living in $Q_b$. The key difference is that the effective theory includes gravity, so we are only allowed to use Neumann boundary conditions to anchor HRT surfaces.  

We work with global coordinates where the metric and dilaton are given by (\ref{eq:Sol to dS JT EOM}). Meanwhile, the bulk metric is given by (\ref{eq:main metric}). The HRT formula in this case tells us that the relevant solutions satisfying Neumann boundary conditions are constant-$\tau$ constant-$\alpha$ surfaces. Therefore, for a spacelike interval $A$ with endpoints $P_1=(\tau_1,\alpha_1)$ and $P_2=(\tau_2,\alpha_2)$ the associated entanglement entropy is given by the sum of the contributions from surfaces ending on the two endpoints,
\begin{align}
S_A&=\sum_{i=1,2}\frac{(\rho_c+\delta\rho_c)-(\rho_b+\delta\rho_b)}{4G_N}\bigg|_{P_i}=\sum_{i=1,2}\frac{\varphi_0+\varphi}{4G_N}\bigg|_{P_i}\nonumber\\
&=\frac{2\varphi_0+\varphi_r\,(\sin\alpha_1\,\cosh\tau_1+\sin\alpha_2\,\cosh\tau_2)}{4G_N}\,.
\end{align}
Further, extremizing with respect to $P_1$ and $P_2$ we are left with a `natural' observer that has access to a full static patch so that the endpoints $P_1$ and $P_2$ correspond to the location of its cosmological horizons: $\tau_1=\tau_2=0$ and $\alpha_2=\alpha_1+\pi$. In this case, the entanglement entropy yields the standard Gibbons-Hawking entropy \cite{Cotler:2019nbi}
\begin{equation}\label{eq:lift degeneracy}
   S_{A}=S_{\rm GH}=\tfrac{\varphi_0}{2G_N}\,.
\end{equation}
where $\varphi_0$ is the topological term of the dilaton (\ref{eq:phi0}).

\subsubsection{Holographic complexity}\label{eq:Holo complexity dilaton}
We will now consider the holographic complexity of the effective dS JT theory. A similar calculation was recently done \cite{Bhattacharya:2023drv} for the case of standard AdS wedge holography, however, here we will extend their results to the dS case. For the purposes of this section, we may keep both the UV and IR branes in the bulk, as in the previous subsection, or we may take $Q_c$ to the boundary, as done in Sec. \ref{sec:QI}.

\subsubsection*{CV proposal}
\begin{figure}[t!]
    \centering
\includegraphics[width=5cm]{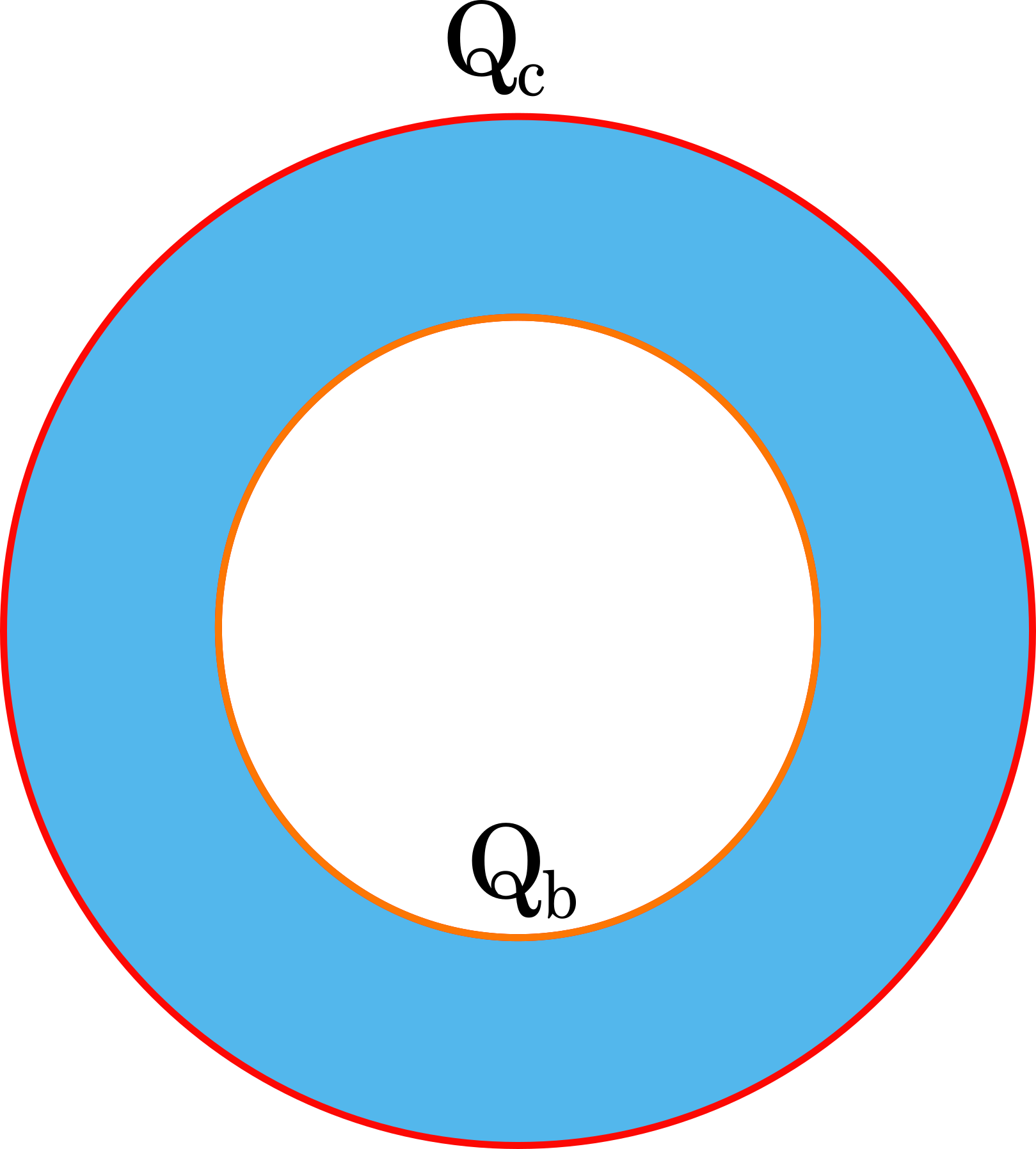}
\hspace{2.5cm}
\includegraphics[width=5cm]{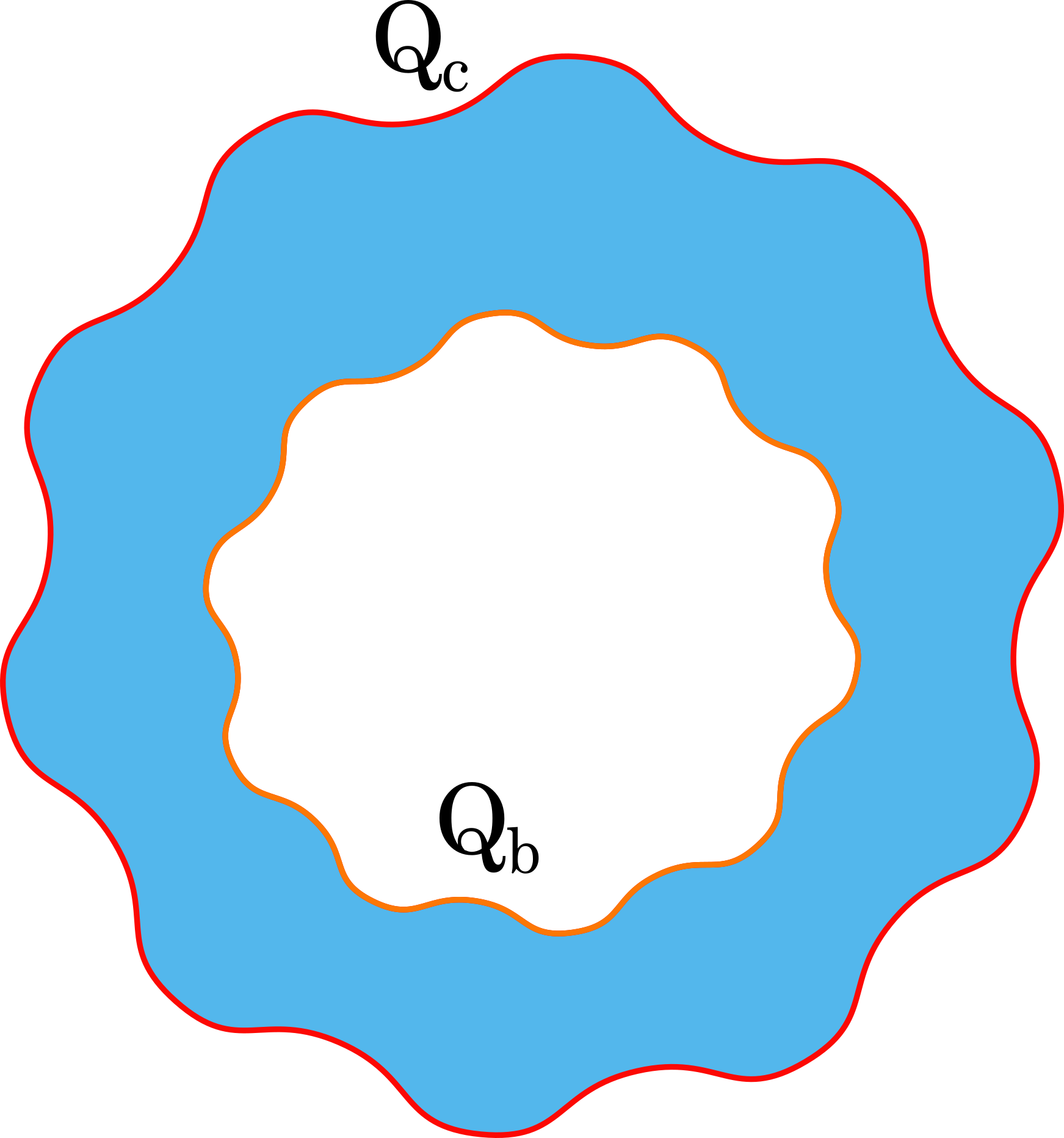}
    \caption{Maximal volume enclosed by the two dS branes corresponds to the complexity of the state. We depict the volume enclosed by two rigid branes (left panel) and two fluctuating branes (right panel) on a fixed global time slice. Without loss of generality, we could also send $Q_c$ to the AdS boundary, in which case its fluctuations would be suppressed. In this limit, $Q_c$ would act as a non-gravitating bath and $Q_b$ would be effectively described by 2d dS JT gravity (+ higher corrections).}
    \label{fig:comps}
\end{figure}

We begin by considering CV duality ---see Fig. \ref{fig:comps} for reference.
The evaluation in this case is just as in (\ref{eq:Vol d}) for $d=2$, but now, we allow for brane fluctuations (\ref{eq:brane fluctuations}),
\begin{equation}\label{eq:compl fluctuate}
\begin{aligned}
    {\mathcal{V}}&=2\int_{\sigma_{\rm st}}^{r_t} \rmd \sigma\sqrt{-f(r)\,\dot{v}(\sigma)^2+2\dot{v}(\sigma)\dot{r}(\sigma)}\int^{\rho_c+\delta\rho_c}_{\rho_b+\delta\rho_b}\rmd\rho\,\sinh\rho~,\\
\end{aligned}
\end{equation}
where $\sigma_{\rm st}$ indicates the location of the stretched horizon. We work in the limit when $\rho_b,\, \rho_c\gg 1$ and the two branes are close to each other, $\varphi\ll\varphi_0\ll1$, in which case, the higher-order corrections to the effective theory are negligible. Further, in this limit, the volume computation simplifies because the above conditions ensure that the embedding functions for the maximal volume surfaces do not depend on the bulk radial direction $\rho$.

Hence, to linear order in $\varphi_0$ and $\varphi$, the volume (\ref{eq:compl fluctuate}) becomes:
\begin{equation}
     \begin{aligned}
    {\mathcal{V}}&\approx\sinh\rho_b\int_{\mathcal{B}}\rmd \sigma\sqrt{-f(r)\,\dot{v}(\sigma)^2+2\dot{v}(\sigma)\dot{r}(\sigma)}\,(\varphi_0+\varphi(\sigma))~.
     \end{aligned}
\end{equation}
Note that $\sinh{\rho_b}$ is the effective curvature scale on the IR brane. Then, as in \cite{Bhattacharya:2023drv}, we confirm that for pure JT gravity:
\begin{equation}\label{eq:Vol total}
    {\mathcal{V}}=\sinh\rho_b\int_{\mathcal{B}}\rmd\sigma\sqrt{h}\,(\varphi_0+\varphi(\sigma))~.
\end{equation}
That is, instead of getting only the volume of the induced geometry on the brane, we get an additional factor $\varphi_0+\varphi(\sigma)$ multiplying the induced metric on the brane in (\ref{eq:Vol total}). This is expected as this factor grows proportionally to the volume of the transverse space to the brane and should enter into the extremization process in order to find the holographic complexity.\footnote{Likewise when JT gravity is derived for a dimensional reduction of a higher dimensional near-extremal black hole, the factor $\varphi_0+\varphi(\sigma)$ should appear as it captures the volume of the transverse sphere away from extremality. This modification of CV was indeed found in this way for dS JT gravity \cite{Anegawa:2023wrk} where it was dubbed the ``refined volume'' conjecture.} To analyze the result (\ref{eq:Vol total}), let us denote the total dilaton in the static patch as
\begin{equation}\label{eq:Total dilaton}
    \Phi(r)\equiv\varphi_0+\varphi(r)\quad\text{where}\quad\,\,\,\varphi(r)=\varphi_h r\,,
\end{equation}
and $\varphi_h\ll 1\in\mathbb{R}$. Then, the modifications to the volume complexity (\ref{eq:Vol d}) and its time evolution can be found to be
\begin{equation}\label{eq:main volume CV}
\begin{aligned}
    t&=-2\int_{r_{\rm st}}^{r_t}\tfrac{\rmd r\,P_v}{f(r)\sqrt{P_v^2+f(r)\Phi(r)^2}}~,\\
    {\mathcal{V}}&=2\sinh\rho_b\int_{r_{\rm st}}^{r_t}\tfrac{\rmd r\,\Phi(r)^2}{\sqrt{P_v^2+f(r)\Phi(r)^2}}~,
\end{aligned}
\end{equation}
where the turning point $r_t$ is determined from the equation $P_v^2=(r_t^2-1)\Phi(r_t)^2$. One can further manipulate (\ref{eq:main volume CV}) to obtain
\begin{equation}\label{eq:manipulation CV dilaton}
    {\mathcal{V}}={\sinh\rho_b}\bigg(P_v\, t+2\int_{r_{\rm st}}^{r_t}\rmd r\frac{\sqrt{P_v^2+f(r)\Phi(r)^2}}{f(r)}\bigg)~.
\end{equation}
From (\ref{eq:main volume CV}) and (\ref{eq:manipulation CV dilaton}) we can obtain the so-called hyperfast growth for the modified CV prescription, namely that
\begin{equation}
\begin{aligned}
    \!\!\!\lim_{P_v\rightarrow\infty}\!t&=2\,\text{arctanh}\,r_{\rm st}\equiv t_\infty~,\\
    \!\!\!\lim_{P_v\rightarrow\infty}\!\dv{\mathcal{V}}{t}&={\sinh\rho_b}\, P_v\approx \sinh{\rho_b}\qty[\varphi_h \qty(\frac{\alpha}{(t_\infty-t)^2}+\frac{\beta}{t_\infty-t})+\varphi_0\frac{\gamma}{\sqrt{t_\infty-t}}+\dots],
\end{aligned}
\end{equation}
where $\alpha$, $\beta$, $\gamma$ are constant numbers and the ellipsis denotes terms that remain finite as $t\to t_\infty$. We note that the critical time $t_\infty$ remains unaffected in comparison to standard CV, however, the divergence increases in order. This is in fact expected, because the fluctuations eventually dominate as $r$ grows (\ref{eq:Total dilaton}), even though $\varphi_h$ is small. Other than that, the modified CV prescription in our wedge holographic model correctly reproduces the qualitative behavior obtained for complexity in the half-reduction dS JT gravity model \cite{Anegawa:2023wrk}. As we remarked in Sec \ref{Sec:CV rigid}, the time evolution for $t>t_\infty$ requires analytic continuation and holographic complexity can have different behaviors, such as the late-time growth proposal in \cite{Jorstad:2022mls}.

\subsubsection*{`Complexity=Anything' proposal}
Next, we can repeat an exactly parallel argument for codimension-one CAny proposals, now with the inclusion of the total dilaton (\ref{eq:Total dilaton}). The resulting contributions for the complexity growth (\ref{eq:C epsilon inter}) in $d=2$ to linear order in the fluctuations become
\begin{equation}\label{eq:New dilaton complexity}
\begin{aligned}
    \mathcal{C}^\epsilon &=\tfrac{\sinh\rho_b}{G_N}\int_{\Sigma_\epsilon}\rmd\sigma\,\sqrt{-f(r)\dot{v}^2+2\dot{v}\dot{r}}\,\Phi(r)\,,\\
    \mathcal{C}_{\rm CMC}&=\tfrac{\sinh\rho_b}{G_N}\sum_{\epsilon}\alpha_\epsilon\int_{\Sigma_\epsilon}
    \rmd\sigma\,\qty[\sqrt{-f(r)\dot{v}^2+2\dot{v}\dot{r}}-\epsilon{\,\sinh\rho_b\,K_\epsilon}\dot{v}r]\Phi(r)~,
\end{aligned}
\end{equation}
which can be covariantly expressed (respect to (\ref{eq:main metric})) as
\begin{equation}\label{eq:CAny dilaton}
\begin{aligned}
    \mathcal{C}^\epsilon &= \tfrac{\sinh\rho_b}{G_N}\int_{\Sigma_\epsilon}\rmd\sigma\,\sqrt{h}~\Phi(\sigma)~,\\
    \mathcal{C}_{\rm CMC}&=\tfrac{\sinh\rho_b}{G_N}\biggl[\sum_{\epsilon}\alpha_\epsilon\int_{\Sigma_\epsilon}\rmd\sigma\,\sqrt{h}\Phi(\sigma)+\alpha_B\sinh\rho_b\int_{\mathcal{M}}\rmd^{2}x\sqrt{-g}\,\Phi(x)\biggr]~.
\end{aligned}
\end{equation}
where $\Phi(x)$ represents the (possibly time-dependent) total dilaton. We refer the reader to App. \ref{Sec:Dynamical CAny} for the evaluation of (\ref{eq:CAny dilaton}) for a time-independent dilaton field $\Phi(r)$ in (\ref{eq:Total dilaton}). In this case, we recover a late-time growth, similar to (\ref{eq:Vol late times}), as:
\begin{equation}\label{eq:Late times CAny dilaton}
    \lim_{t\rightarrow\infty}\dv{t} \mathcal{C}^\epsilon =\frac{\sinh\rho_b}{G_N}\sqrt{-f(r_{f})}\Phi(r_{f})^2\,,
\end{equation}
with (\ref{eq:rt location}) respected. This modifies (\ref{eq:rt location}) to:
\begin{equation}\label{eq:Phi r_f}
   (\Phi(r_f)f'(r_f)+2f(r_f)\Phi'(r_f))^2-4f(r_f)\sinh^2\rho_bK_\epsilon^2(\Phi(r_f)+r\Phi'(r_f))=0~.
\end{equation}
Hence, for the modified CAny prescription, the bounds on $K_{\epsilon}$ (\ref{eq:Key to crit}) determining the late-time behavior of complexity turn out to depend on $\Phi(r)$ in a non-trivial manner.

\section{Discussion and outlook}\label{Sec:Conclusion}
To summarize, we have constructed a new setting that describes a pair of coupled and entangled uniformly accelerated universes in the context of braneworld holography and use it to study information-theoretic aspects of dS space from the AdS bulk spacetime. In this setting, a pair of branes are connected through codimension-2 Euclidean defects that are timelike separated. We coined this configuration the dS wedge holography proposal. This system can be seen as a Randall-Sundrum type-II scenario \cite{Karch:2000ct} with two dS ETW branes \cite{Chacko:2013dra,Karch:2020iit,Mishra:2022fic}. Using this construction, we explored the notions of holographic entanglement entropy and holographic complexity. 

For most of our work, we assumed that the UV brane is placed very close to or exactly at the asymptotic AdS boundary while keeping the IR brane at some fixed radial distance. The system thus describes a dynamical dS universe coupled to a non-gravitating bath, represented in this case by a dS QFT with a mass gap. Furthermore, the two theories are in an entangled state in the full Hilbert space $\mathcal{H}_{\text{UV}}\otimes\mathcal{H}_{\text{IR}}$, as is evident from the emergent space connecting the branes through the bulk. Indeed, as explained in Sec.~\ref{Sec:setting}, we can interpret our setting as a tunneling instanton describing membrane pair creation \cite{Garriga:1993fh,Arcos:2022icf}, analogous to the Schwinger effect (which gives rise to entangled EPR pairs) \cite{Frob:2014zka,Fischler:2014ama}.

Our first application was the calculation of the entanglement entropy of subregions for an observer collecting radiation in the UV theory. Via the RT prescription, the calculation required studying extremal area surfaces in the ambient AdS space, anchored in the UV but possibly connecting with the IR brane. The result of this calculation was a Page curve transition (provided the two branes are separated beyond a certain threshold) for the boundary observer, with the IR brane playing a crucial role in the restoration of unitary. The appearance of the island phase was a result of UV/IR entanglement and could be explained as the result of confinement in the UV CFT, or more specifically, due to the mass gap \cite{Klebanov:2007ws,Jokela:2020wgs}. See \cite{Emparan:2023dxm,VanRaamsdonk:2021qgv} for further discussion on confinement in the framework of braneworlds. Previous attempts in the context of Randal-Sundrum braneworld models \cite{Cohen:1998zx} with one brane required the introduction of an IR regulator that depended on UV physics in order for the Bekenstein bound in dS space to be obeyed. In our setting, the IR regulator is replaced by the location of the IR brane. We derived an equivalent inequality resulting from the Page transition to hold that directly relates to such IR-UV dependence.

Our second application was the calculation of
holographic complexity within the framework of CV duality and the `complexity=anything' proposals. We worked in the context of static patch holography, in which case the relevant surfaces are meant to be anchored along stretched horizons in the dS space. For a set of complexity proposals, including the CV, the resulting evolution was shown to display the hyperfast growth phenomenon, i.e. a diverging growth of complexity at finite time scales. We illustrated this effect in detail for the CV proposal. On the other hand, late-time growth can be achieved for a different set by properly adjusting the parameters in the types of proposals we studied, reminiscing of the results of  \cite{Aguilar-Gutierrez:2023zqm}. See \cite{ACHKKS} for upcoming work on holographic complexity in the presence of ETW branes and \cite{Bhattacharya:2021jrn} for similar work in the context of subregion complexity.

We then studied a refined version of our model that includes fluctuations transverse to the branes. When specializing to a three-dimensional bulk, the resulting effective description was shown to be pure dS JT gravity + higher order corrections that could be suppressed by placing the branes sufficiently close to each other and taking the limit of small fluctuations, or by considering a $\mathbb{Z}_2$-symmetric version of the wedge setup. We investigated how the fluctuations would enter in the calculation of holographic entanglement entropy and complexity in the CAny proposals, and interpreted these calculations from the effective brane description. We obtained corrections to such quantities given by dilaton factors that are nonetheless very important in matching the qualitative behavior of complexity for dS JT gravity that arises from the CA proposal with the CV \cite{Anegawa:2023wrk}. Our calculations can further be seen as a `derivation' of CAny proposals for dS JT gravity, in the spirit of \cite{Bhattacharya:2023drv}.

A natural extension to our results would be to do the numerical analysis of entanglement and complexity in the higher dimensions, which we mostly omitted to retain analytical control. Another possibility is further expanding the analysis with the inclusion of an accelerating black hole in the bulk allowing for the dS space branes to support a black hole as well. In particular, this configuration in the four bulk dimensions has been used to develop the quantum SdS$_3$ black hole \cite{Emparan:2022ijy, Panella:2023lsi}. The main advantage is the exact analytic control in the quantum backreaction on the brane effective description. It would be very interesting to study if the Page curve and analysis of complexity could be performed within such a setting. In particular, \cite{Emparan:2021hyr} has argued that there are semiclassical corrections in the case of a quantum BTZ black hole. Perhaps a similar treatment combining the lessons explored in this manuscript could be performed. In particular, it would be interesting to learn about such semiclassical corrections for the type of CAny proposals investigated in our work. 

Likewise, the effects of semiclassical matter corrections can be studied in 2d dS JT gravity, by considering three-dimensional bulk geometries beyond pure AdS$_3$. In particular, there is a suggestion in \cite{Carrasco:2023fcj} that quantum corrections to volume complexity, coming from the addition of a Polyakov term to the pure JT gravity lead to a modification of CV that includes a `bulk complexity' term. It would be interesting to recover this effect in some extensions of our setting with excited states.  In particular, quantum matter on the branes naturally appears when we integrate out normalizable modes in the bulk spacetime \cite{Emparan:1999wa,Emparan:1999fd}. In that case, corrections to holographic complexity due to quantum backreaction should appear, and it would be interesting to compare with the proposal in \cite{Carrasco:2023fcj}. A possible extension of such arguments to the CAny codimension-1 proposals is worth considering as well.

\begin{figure}
    \centering
\includegraphics[width=3.5cm]{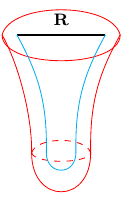}
\hspace{0.5cm}
\includegraphics[width=6cm,trim={0 -3mm 0 0},clip]{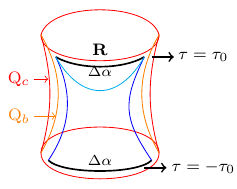}
\vspace{-4mm}
    \caption{Left: Holographic pseudo entropy for dS$_2$ space in the Hartle-Hawking no-boundary proposal with timelike and spacelike geodesics shown in cyan. Right: Holographic pseudo entropy for dS$_2$ branes embedded on AdS$_3$ space. The HRT surfaces for the connected (blue) and disconnected (cyan) phases between timelike separated codimension-2 defects. The entangling region $\mathbf{R}$ is shown in black.}
    \label{fig:TEE phases}
\end{figure}

Finally, there has been a lot of interest in new notions of holographic entanglement entropy in the context of dS/CFT holography \cite{Strominger:2001pn}. Among them is the quantum information theoretic notion of holographic pseudo entropy \cite{Nishioka:2021cxe,Mollabashi:2021xsd,Chen:2023gnh,Doi:2022iyj,Narayan:2023ebn}. It quantifies the entanglement entropy of a spatial subregion, $\mathbf{R}$, of a future infinity in dS space, see Fig.~\ref{fig:TEE phases} (Left). In the Hartle-Hawking construction of the dS space prepared from an Euclidean sphere, the pseudo entropy is a complex-valued function. The real part corresponds to the spacelike interval in the sphere instanton, while the imaginary part is for a timelike interval in dS space. See \cite{Chen:2023gnh,Chen:2023prz,Chu:2023zah,Jiang:2023loq,Narayan:2023ebn,Jiang:2023ffu,Doi:2023zaf,Doi:2022iyj} for recent discussion about its analytical continuation in AdS/CFT context to a different notion, named timelike entanglement entropy.
 Our wedge holographic proposal might provide a new test ground for such types of generalized notions of holographic entanglement entropy in dS space. As shown in Fig.~\ref{fig:manybranes} there is a region at $\tau_g=\pm \pi/2$ where the branes intercept, which we denote as the codimension-2 defects. Since the defects have a timelike separation, one can directly obtain the pseudo entropy holographically from geodesics in the bulk, see Fig \ref{fig:TEE phases} (Right). This evaluation is proportional to the holographic pseudo entropy of dS$_2$ space when there are only timelike geodesics. The details are shown in the App. \ref{App:timelike}. The explicit evaluation assumes a fixed brane tension. We also expressed what modifications must be made when the scalar mode of fluctuations is incorporated. It would be interesting to study these effects in detail from the perspective of the CFT dual in the defects.

\textbf{Note added}: While this work was near completion, \cite{Chang:2023gkt} appeared on the ArXiv, which has a similar flavor to our work, in particular to Sec.~\ref{Sec:Dilaton HEE}. A key difference between our approaches is that \cite{Chang:2023gkt} considers dS JT gravity as the intrinsic theory on the IR brane, while we derive dS JT as an effective description due to fluctuations.

\section*{Acknowledgements}
We would like to thank Stefano Baiguera, José Barbón, Roberto Emparan, Hao Geng, Filip Landgren, Andrea Legramandi, Dominik Neuenfeld, Andrew Svesko, and Watse Sybesma for several discussions and useful correspondence, and  Pratik Nandy and Martin Sasieta for collaboration in the early stages of this project. The work of SEAG is partially supported by the FWO Research Project G0H9318N and the inter-university project iBOF/21/084. AKP and JFP are supported by the ‘Atracci\'on de Talento’ program (Comunidad de Madrid) grant 2020-T1/TIC-20495, by the Spanish Research Agency via grants CEX2020-001007-S and PID2021-123017NB-I00, funded by MCIN/AEI/10.13039/501100011033, and by ERDF A way of making Europe. SEAG further acknowledges the Instituto de Física Teórica UAM/CSIC, the University of Amsterdam, and the Delta Institute for Theoretical Physics for their hospitality and support during various stages of the project. 

\appendix

\section{Coordinate systems}\label{App:Coord}
For the convenience of the reader, we include a list of the coordinate systems employed in the derivation of the entanglement entropy in Sec. \ref{Sec:Holo EE}, for arbitrary dimensions:
\begin{itemize}
\item AdS$_{d+1}$ space in global coordinates can be expressed as:
\begin{equation}\label{eq:global coordinates}
\rmd s^2=\rmd \rho_g^2-\cosh^2\rho_g\,\rmd\tau_g^2+\sinh^2\rho_g(\rmd\alpha^2+\cos^2\alpha\,\rmd\Omega_{d-2}^2)\,.
\end{equation}
The area functional (\ref{eq:Area functional}) to compute holographic entanglement entropy of a disk-shaped region
is: 
\begin{equation}\label{eq:Sgen global time}
\mathcal{A}=\Omega_{d-2}\int\rmd s~(\sinh\rho_g\cos\alpha)^{d-2}\sqrt{{\rho_g'}^2-\cosh^2\rho_g{\tau_g'}^2+\sinh^2\rho_g\,{\alpha'}^2}~,
\end{equation}
where the prime denotes derivatives with respect to the affine parameter $s$, and
\begin{equation}
    \Omega_{d-2}=\frac{2 \pi ^{\frac{d-1}{2}}}{\Gamma
   \left(\frac{d-1}{2}\right)}~.
\end{equation}
    \item AdS$_{d+1}$ space foliated with (global) dS$_d$ slices can be expressed as:
    \begin{equation}\label{eq: AdS with global dS foliation}
\rmd s^2=\rmd \rho^2+\sinh^2\rho [-\rmd\tau^2+\cosh^2\tau~(\rmd\alpha+\cos^2\alpha\,\rmd\Omega_{d-2}^2)]\,.
\end{equation}
The slices with $\rho=\rho_{b,c}$ are shown in Fig.~\ref{fig:manybranes}, which can be fixed by suitably picking the brane tensions. The area functional (\ref{eq:Area functional}) in this coordinate system is:
\begin{equation}\label{eq:entropy dS global time slices}
    \mathcal{A}=\Omega_{d-2}\int\rmd s\,\qty(\sinh\rho\,\cos\alpha\,\cosh\tau)^{d-2}\sqrt{\rho'^2-\tfrac{\sinh^2\rho}{\sinh^2\rho_b}\tau'^2+\tfrac{\cosh^2\tau\,\sinh^2\rho}{\sinh^2\rho_b}\alpha'^2}\,,
\end{equation}

 \item The map between two previous coordinate systems is given by:
\begin{equation}
\begin{aligned}\label{eq:map global to dS foliation}
    \tan\tau_{g}&=\sinh\tau\tanh\rho~,\\
\sinh{\rho_{g}}&=\sinh{\rho}\cosh{\tau}~.
\end{aligned}
\end{equation}
\end{itemize}

\section{Details on the evaluation of the `Complexity=Anything' proposal}\label{App:CAny details}
\subsection{Rigid branes}
Therefore we proceed in the analysis considering only extremal surfaces of the form $v(\sigma)$, $r(\sigma)$ in the regime $\rho_c,\,\rho_b\gg1$, where we can integrate out $\rho$, as in (\ref{eq:ftilde}). We obtain
\begin{equation}\label{eq:C epsilon inter}
\begin{aligned}
    \mathcal{C}^\epsilon &=\tfrac{\Omega_{d-2}}{G_N}I_{(d-1)}(\rho_c,\,\rho_b)\int_{\Sigma_\epsilon}\rmd\sigma\,r^{d-2}\sqrt{-f(r)\dot{v}^2+2\dot{v}\dot{r}}\,\,,\\
    \mathcal{C}_{\rm CMC}&=\tfrac{\Omega_{d-2}}{G_N}I_{(d-1)}(\rho_c,\,\rho_b)\sum_{\epsilon}\alpha_\epsilon\int_{\Sigma_\epsilon}
    \rmd\sigma\,\mathcal{L}_\epsilon~,
\end{aligned}
\end{equation}
where
\begin{align}
    &\mathcal{L}_\epsilon\equiv r^{d-2}\sqrt{-f(r)\dot{v}^2+2\dot{v}\dot{r}}-\tfrac{k^\epsilon_{d}(\rho_c,\,\rho_b)}{d-1}\dot{v}r^{d-1}~,\\
    &k^\epsilon_{d}(\rho_c,\,\rho_b)\equiv\frac{\epsilon K_\epsilon I_{(d)}(\rho_c,\,\rho_b)}{I_{(d-1)}(\rho_c,\,\rho_b)}~;\label{eq:k^epsilon_d}
\end{align}
$\sigma$ is the parametrization of the coordinates $v(\sigma)$, $r(\sigma)$ on the slice $\Sigma_\epsilon$. For the choice
\begin{equation}\label{eq:gauge choice}
\sqrt{-f(r)\dot{v}^2+2\dot{v}\dot{r}}=r^{d-2}~.
\end{equation}
The equation of motion corresponding to (\ref{eq:C epsilon inter}) can be expressed as
\begin{equation}\label{eq:potential}
    \dot{r}^2+\mathcal{U}(P_v^\epsilon,\,r)=0~,
\end{equation}
where
\begin{align}
P_v^{\epsilon}&\equiv\pdv{\mathcal{L}_\epsilon}{\dot{v}}=\dot{r}-\dot{v}\,f(r)-\tfrac{k^\epsilon_{d}(\rho_c,\,\rho_b)}{d-1}\,r^{d-1}~;\\
    \mathcal{U}(P_v^{\epsilon},\,r)&\equiv -f(r) r^{2(d-2)} - \left(P_v^\epsilon+ \tfrac{k^\epsilon_{d}(\rho_c,\,\rho_b)}{d-1}  r^{d-1}\right)^2~.\label{eq:potential arbitrary}
\end{align}
Then, we can express (\ref{eq:C epsilon inter}) as
\begin{equation}\label{eq:Cepsilon car}
    \mathcal{C}^\epsilon=\tfrac{2\Omega_{d-2}}{G_N}I_{(d-1)}(\rho_c,\,\rho_b)\int_{r_{\rm st}}^{r_{t}}\tfrac{r^{2(d-1)}}{\sqrt{-\mathcal{U}(P_v^\epsilon,\,r)}}\rmd r~.
\end{equation}
In a similar way, the parameter $t$ can be expressed as
\begin{equation}\label{eq:time def}
\begin{aligned}
t&=\int_{\Sigma_\epsilon} \rmd r\tfrac{\dot{t}}{\dot{r}}=\int_{\Sigma_\epsilon} \rmd r\tfrac{\dot{v}-\dot{r}/{f(r)}}{\sqrt{-\mathcal{U}(P_v^\epsilon,\,r)}}\\
    &= -2\int_{r_{\rm st}}^{r_{t}} \tfrac{\rmd r}{f(r)\,\sqrt{-\mathcal{U}(P_v^\epsilon, r)}} \left(P_v^\epsilon + \tfrac{k^\epsilon_{d}(\rho_c,\,\rho_b)}{d-1} r^{d-1} \right) ~.
\end{aligned}
\end{equation}
We proceed to evaluate (\ref{eq:Cepsilon car}) with (\ref{eq:time def}) carefully. Since $U(P_v^\epsilon,\,r_{t})=0$ by definition, we need to take care of the denominator in (\ref{eq:Cepsilon car}, \ref{eq:time def}) at each of the turning points. We do so by adding a subtracting a term:
\begin{align}\label{eq:Cepsilon close form}
    \frac{\mathcal{C}^\epsilon}{I_{(d-1)}(\rho_c,\,\rho_b)}=&-\tfrac{2\Omega_{d-2}}{G_N}\sqrt{-f(r_t)r_t^{2(d-2)}}\int_{r_{\rm st}}^{r_{t}}\tfrac{\left(P_v^\epsilon + \frac{k^\epsilon_{d}(\rho_c,\,\rho_b)}{d-1} r^{d-1} \right)\rmd r}{f(r)\sqrt{-\mathcal{U}(P_v^\epsilon, r)}}\\
    &+\tfrac{2\Omega_{d-2}}{G_N}\int_{r_{\rm st}}^{r_{t}}\tfrac{f(r)\,r^{2(d-2)}+\sqrt{-f(r_{t})r_{t}^{2(d-2)}}\left(P_v^\epsilon + \frac{k^\epsilon_{d}(\rho_c,\,\rho_b)}{d-1} r^{d-1} \right)}{f(r)\sqrt{-\mathcal{U}(P_v^\epsilon,\,r)}}~.\nonumber
\end{align}
Then, we can identify the relationship between time in (\ref{eq:time def}) and complexity in (\ref{eq:Cepsilon close form})
\begin{equation}
\begin{aligned}
    \tfrac{G_N}{\Omega_{d-2}}\tfrac{\mathcal{C}^\epsilon}{I_{(d-1)}(\rho_c,\,\rho_b)}=&\sqrt{-f(r_{t})r_{t}^{2(d-2)}}t\\
    &+2\int_{r_{\rm st}}^{r_{t}}\tfrac{f(r)\,r^{2(d-2)}+\sqrt{-f(r_{t})r_{t}^{2(d-2)}}\left(P_v^\epsilon + \frac{k^\epsilon_{d}(\rho_c,\,\rho_b)}{d-1} r^{d-1} \right)}{f(r)\sqrt{-\mathcal{U}(P_v^\epsilon,\,r)}}~.
\end{aligned}
\end{equation}
We can then straightforwardly take the time derivative,
\begin{align}
    \frac{\dv{\mathcal{C}^\epsilon}{t}}{I_{(d-1)}(\rho_c,\,\rho_b)}=&\tfrac{\Omega_{d-2}}{G_N}\sqrt{-f(r_{t})r_{t}^{2(d-2)}}\label{eq:complexity growth}\\
    &+\tfrac{2\Omega_{d-2}}{G_N}\dv{P_v^\epsilon}{t}\int_{r_{\rm st}}^{r_{t}}\rmd r\tfrac{
        r^{2(d-2)} \left(\sqrt{-f(r_{t}) r_{t}^{2(d-2)}} 
        - \qty(P_v^\epsilon + \tfrac{k_{d}(\rho_c,\,\rho_b)}{d-1} r^{d-1})\right)}{\qty(-\mathcal{U}(P_v^\epsilon,\,r))^{3/2}}~.\nonumber
\end{align}
In the late-time limit, we then recover (\ref{eq:Vol late times}).

\subsection{Dynamical branes}\label{Sec:Dynamical CAny}
In $d=2$, the gauge choice (\ref{eq:gauge choice}) is modified to
\begin{equation}
\sqrt{-f(r)\dot{v}^2+2\dot{v}\dot{r}}=\Phi(r)~,
\end{equation}
(\ref{eq:potential arbitrary}) is then modified as:
\begin{equation}
\begin{aligned}
    P_v^\epsilon&=\dot{r}-f(r)\dot{v}-\epsilon ~\sinh\rho_b~K_{\epsilon}~r~\Phi(r)~,\\
    \mathcal{U}(r,\,P_v^\epsilon)&=-f(r)\Phi(r)^2-(P_v^\epsilon+\epsilon~\sinh\rho_b~ K_\epsilon ~r~\Phi(r))^2~,
\end{aligned}
\end{equation}
and similarly for (\ref{eq:Cepsilon car}) and (\ref{eq:time def})
\begin{equation}
    \mathcal{C}^\epsilon=\tfrac{2\sinh\rho_b}{G_N}\int_{r_{\rm st}}^{r_{t}}\tfrac{\Phi(r)^2}{\sqrt{-\mathcal{U}(P_v^\epsilon,\,r)}}\rmd r~,
\end{equation}
\begin{equation}
    t= -2\int_{r_{\rm st}}^{r_{t}} \tfrac{\left(P_v^\epsilon +\epsilon \sinh\rho_b{K_\epsilon} r\Phi(r) \right)\rmd r}{f(r)\,\sqrt{-\mathcal{U}(P_v^\epsilon, r)}}
\end{equation}
respectively. In that case, (\ref{eq:Cepsilon close form}) is now express as:
\begin{align}
    \frac{G_N\mathcal{C}^\epsilon}{
    \sinh\rho_b}=&-2\sqrt{-f(r_t)}\Phi(r_t)^2\int_{r_{\rm st}}^{r_{t}}\tfrac{\left(P_v^\epsilon +\epsilon\sinh\rho_b K_\epsilon r\Phi(r) \right)\rmd r}{f(r)\sqrt{-\mathcal{U}(P_v^\epsilon, r)}}\\
    &+2\int_{r_{\rm st}}^{r_{t}}\tfrac{f(r)\Phi(r)^2+\sqrt{-f(r_{t})}\Phi(r_{t})^2\left(P_v^\epsilon +\epsilon\sinh\rho_b K_\epsilon r\Phi(r) \right)}{f(r)\sqrt{-\mathcal{U}(P_v^\epsilon,\,r)}}\nonumber\\
    =&\sqrt{-f(r_{t})}\Phi(r_t)^2t\\
    &+2\int_{r_{\rm st}}^{r_{t}}\tfrac{f(r)\Phi(r)^2+\sqrt{-f(r_{t})}\Phi(r_{t})^2\left(P_v^\epsilon +\epsilon\sinh\rho_b K_\epsilon r\Phi(r) \right)}{f(r)\sqrt{-\mathcal{U}(P_v^\epsilon,\,r)}}~,\nonumber
\end{align}
such that (\ref{eq:Late times CAny dilaton}) is found.

\section{Entanglement entropy between the codimension-2 defects}\label{App:timelike}
An alternative notion of entropy in the wedge holographic setting is to consider the entanglement between the two defects \cite{Geng:2020fxl,Akal:2020wfl}, located where the two branes intercept --- see Fig.~\ref{fig:manybranes}. The key difference from the standard wedge holographic settings is that, in the case of dS branes, the defects represent Euclidean CFTs and are timelike separated. Thus, the notion of entanglement that we need would be a version of ``pseudo entropy'' recently introduced in \cite{Nishioka:2021cxe,Mollabashi:2021xsd,Chen:2023gnh,Doi:2022iyj,Narayan:2023ebn} or ``timelike entanglement entropy'' \cite{Chen:2023gnh,Chen:2023prz,Chu:2023zah,Jiang:2023loq,Narayan:2023ebn,Jiang:2023ffu,Doi:2023zaf,Doi:2022iyj}. We argue that, from the braneworld perspective, the holographic dual of this quantity would correspond to the area of an extremal surface connecting subregions in the two defects.\footnote{Further entropy-related notions in dS space have been proposed in \cite{Narayan:2022afv,Narayan:2020nsc,Narayan:2017xca,Narayan:2015vda}.} We provide a proposal to evaluate it for the rigid brane system.

We consider that the branes are very close to each other, $\rho_c=\rho_b+\varphi_0$ for $\varphi_0\ll\rho_b$. Employing (\ref{eq: AdS with global dS foliation}) in $d=2$ with $\rho\approx$ constant, we must evaluate the following functional
    \begin{equation}\label{eq:main functional TEE}
        I=\int\rmd\tau \mathcal{L}(\tau,\,\alpha(\tau))~,\quad\mathcal{L}=\sqrt{(\alpha'(\tau))^2\cosh^2\tau-1}~.
    \end{equation}
which corresponds to a geodesic length of dS$_2$ in global coordinates. Set $P_\alpha=\pdv{\mathcal{L}}{\alpha'(\tau)}$ as the conserved charge. Explicit integration gives
\begin{equation}\label{eq:alpha geo}
    \alpha(\tau)-\alpha_0=\text{arctan}\qty(\tfrac{P_\alpha\sinh\tau}{\sqrt{P_\alpha^2-\cosh^2\tau}})~,
\end{equation}
with $\alpha_0$ an integration constant. Consider the global dS$_2$ time slice $\tau=\tau_0$ near $\mathcal{I}^+$ ($\tau_0\rightarrow\infty$). In this slice, one can employ Dirichlet boundary conditions for $\alpha(\tau)$. Denoting $\Delta\alpha\equiv \alpha(\tau_0)-\alpha_0+\tfrac{\pi}{2}$, we find
\begin{equation}\label{eq:Palpha}
    P_\alpha=\tfrac{\cosh(\tau_0)\cot\Delta\alpha}{\sqrt{\csc^2\Delta\alpha-\cosh^2\tau_0}}~.
\end{equation}
On the other hand, integrating (\ref{eq:main functional TEE}) leads to:
    \begin{equation}\label{eq:STE singular}
        I=\sqrt{\tfrac{1-P_\alpha^2\sech^2\tau_0}{P_\alpha^2-\tanh^2\tau_0}}\text{arctanh}\qty[\tfrac{\sinh\tau_0}{\sqrt{\cosh^2\tau_0-P_\alpha^2}}]~.
    \end{equation}
We now consider different entanglement phases for timelike-separated defects. There are connected and disconnected phases, see Fig \ref{fig:TEE phases}. Explicitly:
\begin{align}
S^{(\rm conn)}=\tfrac{\sinh\rho_b}{2G_3}\int_{-\tau_0}^{\tau_0}\rmd \tau~\mathcal{L}~,\quad S^{(\rm disc)}=\tfrac{\sinh\rho_b}{2G_3}\int_{\tau_t}^{\tau_0}\rmd \tau~\mathcal{L}
\end{align}
where $\tau_t$ is the turning point, and the factor of $2$ arises from $\mathbb{Z}_2$ symmetry.

For the connected phase, the explicit extremization with (\ref{eq:Palpha}) and (\ref{eq:Palpha}) gives:
\begin{equation}\label{eq:time like island}
    S^{(\rm conn)}=\tfrac{\sinh\rho_b}{2G_3}\tfrac{\sqrt{2}\sinh2\tau_0\sin\Delta\alpha}{\sqrt{8\sin^2\Delta\alpha\,\cosh^2\tau_0-7-\cosh4\tau_0}}\text{arctanh}\qty(\tfrac{\sqrt{\cosh^2\tau_0-\csc^2\Delta\alpha}}{\cosh\tau_0})~.
\end{equation}
Meanwhile, for the disconnected phase, the explicit evaluation shows there is no contribution from the turning point in $S^{(\rm disc)}$, as such
\begin{equation}\label{eq:time like disc}
    S^{(\rm disc)}=\tfrac{1}{2}S^{(\rm conn)}~.
\end{equation}
It would seem that the disconnected contribution produces a lower holographic entanglement entropy with respect to the connected one, and would thus dominate, for arbitrary $\Delta \alpha$, $\tau_0$. Next, we provide the $\tau_0\gg1$ expansion,
\begin{equation}\label{eq:Sconn}
    S^{(\rm conn)}=\rmi\tfrac{\sinh\rho_b}{2G_3}\sin\Delta\alpha\,(\tau_0+\log\sin\Delta\alpha)~.
\end{equation}
which is also valid when $\Delta\alpha\rightarrow0$. The result (\ref{eq:Sconn}) is proportional to the imaginary part of pseudo entropy of dS$_3$ \cite{Doi:2022iyj}, as expected since we only account for time like geodesics of dS$_2$ in Fig.~\ref{fig:TEE phases} (b). In particular when $\tau_0\to\infty$ both (\ref{eq:time like island}) and (\ref{eq:time like disc}) would diverge and become purely imaginary. This is in contrast to Fig \ref{fig:TEE phases} (a) where there is a real contribution to the holographic pseudo entropy given by the spacelike intervals on the Euclidean sphere. Our configuration, Fig \ref{fig:TEE phases} (b), only supports timelike geodesics along the dS$_2$ surface. The fact that the result diverges as $\tau_0\to\infty$ is also expected since it has been argued that there are strong backreaction effects causing quantum singularities in such locations \cite{Emparan:2022ijy}. Moreover, there is no degeneracy about the HRT surface in this case, in contrast to the holographic entanglement entropy encountered in the island phase of Sec. \ref{Sec:Holo EE}, as $\alpha(\tau)$ is uniquely determined from the boundary conditions at $\mathcal{I}^+$. If one incorporates in the analysis the fluctuating in the brane locations, we recall that the effect of the scalar mode in geodesics on the dS$_2$ braneworlds (\ref{eq:Vol total}) for $\rho_c=\rho_b+\varphi_0$ is to add the total dilaton, such that (\ref{eq:main functional TEE}) is modified as:
\begin{equation}
    \mathcal{L}=\Phi(\sigma,\,\tau)~\sqrt{(\alpha'(\tau))^2\cosh^2\tau-1}~.
\end{equation}
All in all, our wedge holographic proposal provides a new testing ground for extensions of the holographic entanglement entropy notions in dS space.

\bibliographystyle{JHEP}
\bibliography{references.bib}
\end{document}